\documentclass[twocolumn]{aastex63}

\usepackage{graphicx}	
\usepackage{amsmath}	
\usepackage{amssymb}
\usepackage{bm}
\usepackage{verbatim}
\usepackage{xcolor}
\usepackage{soul,xcolor}
\usepackage{physics}

\newcommand\bfv{\textbf{V}}
\newcommand\bfu{\textbf{U}}

\shorttitle{Forced rotating shear flows}
\shortauthors{S. Ghosh \& B.Mukhopadhyay}

\begin{document}

\setstcolor{red}

\title{Forced linear shear flows with rotation: 
rotating Couette-Poiseuille flow, its stability and astrophysical implications}

\author{Subham Ghosh}
\affiliation{ Department of Physics, Indian Institute of Science,
Bangalore 560012, India}\thanks{subham@iisc.ac.in (SG), bm@iisc.ac.in (BM)}

\author{Banibrata Mukhopadhyay}
\affiliation{ Department of Physics, Indian Institute of Science,
Bangalore 560012, India}



\begin{abstract}

We explore the effect of forcing on the linear shear flow or plane Couette flow, which is also the background flow in 
the very small region of the Keplerian accretion disk. We show that depending 
	on the strength of 
	forcing and boundary conditions suitable for the systems under consideration, 
	the background plane shear flow and, hence, accretion disk velocity profile modifies to parabolic flow, which is 
plane Poiseuille flow or Couette-Poiseuille flow, depending on the frame of reference. In the presence of rotation, plane 
Poiseuille flow becomes unstable at a smaller Reynolds number under pure vertical as well as threedimensional 
perturbations. Hence, while rotation stabilizes plane Couette flow, the same destabilizes plane Poiseuille flow faster and 
forced local accretion disk. Depending on the 
various factors, when local linear shear flow becomes Poiseuille flow in 
the shearing box due to the presence of extra force, the flow becomes unstable even for the Keplerian rotation and hence 
turbulence will pop in there. This helps in resolving a long standing problem of sub-critical transition to turbulence in 
hydrodynamic accretion disks and laboratory plane Couette flow.

\end{abstract}

\keywords{Accretion -- Hydrodynamics -- Compact objects -- Astrophysical fluid dynamics}


\section{Introduction}
\label{sec:Introduction}
Accretion disks are very exotic astrophysical objects. They are formed 
around a denser and  heavier object, mainly in the form of disk due to the accretion of matter from surroundings. We are 
particularly interested in the region where the gravitational force almost balances the centrifugal force. This region is 
called the Keplerian region, and the flow therein is called the Keplerian flow.  This flow is stable 
under linear 
perturbation, and this stability is called Rayleigh stability.

Nevertheless, to explain the observed physical quantities such as temperature, luminosity, etc. based on 
the Keplerian disks
(see, e.g., \citealt{Frank_2002}), the flow therein must be assumed turbulent. Otherwise, there will be a mismatch of
physical quantities, e.g. temperature, of the orders of magnitude
between theory and observations. \cite{Shakura_1972} and \cite{Lynden-Bell_1974} 
then came up with the idea of turbulent viscosity which is responsible for the transport of matter inward 
in accretion disks and hence the physical observables. However, the reason behind the turbulence was not known until  
\cite{Balbus_1991} proposed an idea of instability involving the coupling between the rotation of 
fluid and the weak magnetic field therein following \cite{Velikhov_1959} and \cite{Chandrasekhar_1960}. This 
instability is known as magneto-rotational instability (MRI), which could bring nonlinearity into the 
system and hence turbulence. Later, \cite{Ogilvie_1996} investigated MRI based
on a more complicated analysis.
Although, MRI succeeds greatly in explaining the origin of turbulence in most of the 
hot flows, 
it fails to explain the same in several sites, e.g. protoplanetary disk 
(\citealt{Bai_2017, Bai_2013_ApJ}), 
cataclysmic variables in their low states (\citealt{Gammie_1998, Menou_2000}), the outer part of active galactic nucleus (AGN) 
disks and the underlying dead zone (\citealt{Menou_2001}). MRI gets suppressed in these cases due to very small ionization of the 
matter therein. Apart from these, the systems with huge Reynolds number ($\gtrsim 10^9$), as argued by \cite{Nath_2015}, 
have larger growth rate due to magnetic transient growth than the growth rate due to MRI. 
\cite{bhatia_2016} however showed that even transient energy growth is ceased to occur beyond a certain magnetic 
field in magnetohydrodynamical shear flows. \cite{Pessah_2005} and \cite{Das_2018}, using local 
and global analysis respectively, showed the stabilization of the axisymmetric MRI above a certain magnitude of a toroidal 
component of the magnetic field for compressible and differentially rotating flows. All these publications showed that MRI
is not a generic way to make the Keplerian flow unstable and hence turbulent. As hydrodynamics is generically there, it is 
worth looking for plausible hydrodynamic instability instead.

However, the Keplerian flow is Rayleigh stable and there is a long debate in the literature 
(\citealt{Dubrulle_2005_a, Dubrulle_2005_b, Dauchot_1995, Rudiger_2001, Klahr_2003, Richard_1999, Kim_2000, Mahajan_2008, 
Yecko_2004, Mukhopadhyay_2011NJPh, Mukhopadhyay_2013}) 
regarding the stability of Rayleigh stable flows, especially in the context of accretion disks. The authors put forward 
their efforts to resolve this issue either analytically or with simulation or experimentally. The authors 
like \cite{Balbus_1996} and \cite{Hawley_1999} concluded that the sustained turbulence was not 
possible in the Keplerian flow from hydrodynamics. Nevertheless, other authors, such as \cite{Lesur_2005}
strongly disagreed with it and discussed about the unavailability of the computer resources to resolve the Keplerian regime.
However, with their extrapolated numerical data, they could not produce astrophysically sufficient subcritical turbulent 
transport in the Keplerian flow. There are other authors too who argued for plausible emergence of hydrodynamical 
instability and hence further turbulence, by transient growth in the case of otherwise linearly stable flows 
(e.g. \citealt{Chagelishvili_2003, Tevzadze_2003, Mukhopadhyay_2011NJPh, man_2005, amn_2005, Cantwell_2010}), 
in laboratory experiment (e.g. \citealt{Paoletti_2012}), 
in simulations in case of accretion disks (e.g. \citealt{Avila_2012}). 

We, therefore, look for hydrodynamics that could plausibly give rise to unstable modes when the dynamics of the fluid 
parcel
is studied in a small cubical shearing box (see, e.g., \citealt{man_2005, our_work}, for details) situated at a particular radius 
in the Keplerian disk. We are particularly motivated and 
inspired by our recent results
(\citealt{our_work}), which explored in detail the effect of forcing in the linearly and nonlinearly perturbed
plane shear flows, with and without rotation, which enlightened the issue of the origin of
hydrodynamical turbulence. In fact, there are other works (\citealt{Ioannou_2001, Mukhopadhyay_2013, Nath_2016, Raz20})
considered an extra forcing to be present in the system. However, in the shearing box, the background flow is of linear
shear profile up to first order approximation (see Appendix \ref{sec:der_bg_flow_without_force} and 
\citealt{Balbus_1996}, for details). This linear shear flow is called plane 
Couette flow. As in the accretion disk, the shearing box is situated at a 
particular radius, it will have an angular frequency. We, therefore, have to consider the effect of rotation while we 
describe the motion of the accretion disk fluid parcel 
inside the shearing box. Now, if an extra force is present there in the shearing box, the background flow no longer remains 
to be linear shear, instead becomes quadratic shear flow what we call plane Couette-Poiseuille flow generally. However, with proper transformation, this flow can be 
transformed into plane Poiseuille flow. This flow further will embed 
with rotation in the context of Keplerian flow. Plane Poiseuille flow without rotation
is unstable under linear twodimensional perturbation having critical Reynolds number 5772.22 with critical wave vector 1.02 
(\citealt{Orszag_1971}). Once it is established that the very local flow (inside the box) in the Keplerian region 
with forcing is plane Poiseuille flow with rotation, then we can argue that the flow inside the shearing box 
is unstable. We therefore plan to explore plane Poiseuille flow in the presence of rotational effect.
Although, the effect of rotation 
on the stability of Poiseuille flow was studied by \citealt{Lezius_1976, Alfredsson_1989}, our work is different form them 
in two aspects. First, we extensively study the eigenspectra of plane Poiseuille flow, as well as Couette-Poiseuille flow, with rotation for 
purely vertical perturbations
and threedimensional perturbations. To the best of our knowledge, this study has not been done 
in an extensive manner yet, particularly the effect of rotation on the stability analysis of Couette-Poiseuille 
flow. Although the Poiseuille flow in the presence of rotation has been studied earlier, to our knowledge, its 
application to the stability of accretion flow has never been explored. Apart from that, analysis of eigenspectra for
Poiseuille flow in the presence of rotation has not been performed yet extensively. However see, e.g., 
\citealt{Hains,Cowley,Balakumar,Savenkov,Klotz}, for various explorations of Couette-Poiseuille 
flow over the years. Second, 
the background flow profile that we consider here is different than those already considered in 
previous works (see \citealt{Lezius_1976, Alfredsson_1989}, for details).

The plan of the paper is the following. In \S\ref{sec:Background flow}, we show how the linear shear flow (or plane Couette 
flow) modifies due to the presence of extra force in the system. In a recent work, we assumed that background does
not practically change due to forcing (\citealt{our_work}), here however we explore the change of background and its consequence
in detail. As the background modifies in the 
presence of extra force, 
the domain of the background also modifies depending on the strength of the force. The relevance of the size of the 
new domain is studied in \S\ref{sec:The new domain}. We write the Navier-Stokes equation for the modified background flow 
in the rotating frame, as the primary plan is to the application in
accretion disks, in \S\ref{sec:N-S_eq_rot_frame} and also obtain the corresponding Reynolds number after
nondimensionalizing it in \S\ref{sec:Defining new Reynolds number}. 
The perturbed flow equations have been formulated appropriately in the same section
but in \S\ref{sec:Perturbation analysis} where we recast the Navier-Stokes equation into Orr-Sommerfeld and Squire equations.
Rotating Poiseuille and Couette-Poiseuille flows under purely vertical and 
threedimensional perturbations 
are explored in detail in \S\ref{sec:per-poi} and \S\ref{sec:cou-poi} 
respectively. Also how 
the stability of respective flows depends on the rotational profile is studied in the same sections. In 
\S\ref{sec:Numerical technique}, we describe the accuracy of our numerical 
results based on the technique we have used in this work. In 
\S\ref{sec:Discussion},
we 
compare plane Poiseuille flow with plane Couette flow in the presence of rotation. In the same section we also compare our 
critical parameters with those in literature. We finally conclude in \S\ref{sec:Conclusion} that depending on the boundary 
conditions and the strength of the extra force, there is a deviation in the 
flow from its linear shear nature. Further, rotation makes 
the flow unstable depending on the parameters and hence the flow plausibly becomes turbulent, which we suggest to be the 
hydrodynamical origin of turbulence 
in accretion disks.

\section{Background flow in the presence of force}
\label{sec:Background flow}
\begin{figure}
\includegraphics[width=\columnwidth]{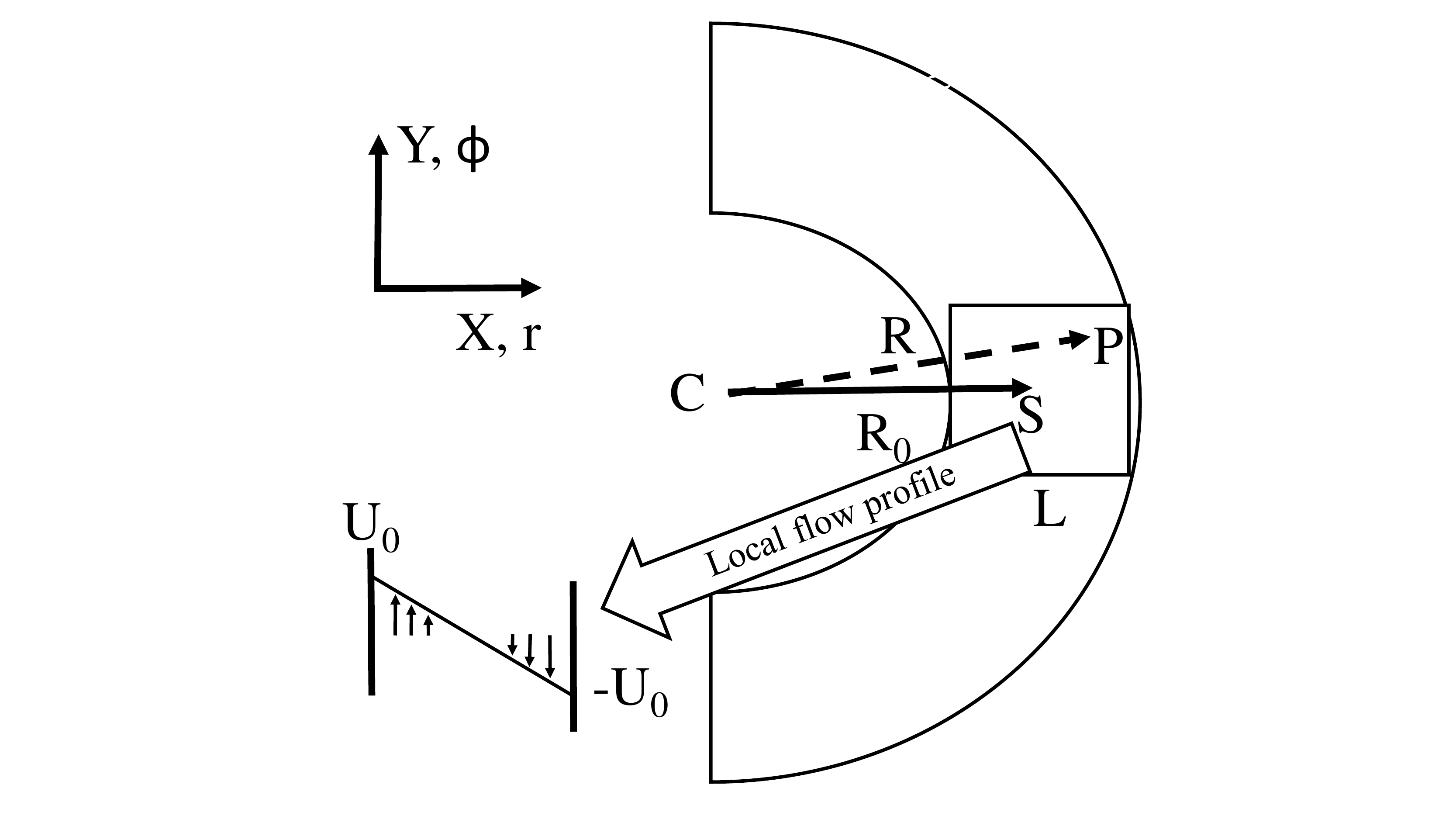} 
  \caption{Schematic diagram of a shearing box centered at the point $S$ inside the small patch of 
accretion disk. The box is of size $L$. $C$ is the center of the accretion disk. $S$ is at a distance $R_0$ from $C$. An 
arbitrary fluid particle inside the box is at $P$ at a distance $R$ from $C$.}
\label{fig:shearing_box_pic} 
\end{figure}

Let us consider a very small cubical box of size $L$ at a particular radius $R_0$ from the center of 
the system as shown in the FIG.~\ref{fig:shearing_box_pic}. At that radius, the box is rotating with 
an angular frequency $\Omega_0$ such that $\Omega=\Omega_0(R/R_0)^{-q}$ and the rotation
parameter $q = 3/2$ for Keplerian flow. In FIG.~\ref{fig:shearing_box_pic}, $S$ is the center of 
the box and the local analysis is done with respect to $S$. See \citealt{man_2005, bhatia_2016, our_work}, for details 
of the reference frame and the background flow therein. Now let us set the local
reference frame or box in such 
a way that the flow, which is along the $\phi$-direction with respect to $C$, to be in the $y$-direction 
and the motion of the either ends of the box in the $x$-direction (in the disk
frame $r$-direction) to be with equal and opposite velocity of magnitude
$U_0$ (see 
fig. 1 of \citealt{our_work}). In that local reference frame
or box, the velocity of Keplerian flow becomes $-q \Omega_0 X$ 
upto the first order approximation. This is the usual background flow 
(see Appendix \ref{sec:der_bg_flow_without_force} and 
also \citealt{Hawley_1995, man_2005, amn_2005, our_work}) in the local region of an accretion disk.
However, due to the presence of external force (may not be random) in 
the flow, the above-mentioned background flow is expected to change. The various possible origins of 
force in the system under consideration, as described earlier by us (\citealt{our_work}) in detail, could be: back 
reactions of outflow/jet to accretion disks, the interaction between the dust grains and fluid parcel in protoplanetary
disks, etc. Using fluid-particle interactions, these possibilities could be modelled in such a way that the extra force 
turns out to be a function of the relative velocity between the fluid and the particles. For details, see section 
2.1 and APPENDIX A of \cite{our_work}. 
In the presence of extra force, let us consider the background flow velocity to be $\bfv$, given by  
\begin{eqnarray}
\bfv= (0, V_Y(X), 0).
\end{eqnarray} 
The corresponding Navier-Stokes equation describing the flow
in the local box is 
\begin{equation}
 \frac{\partial \bfv}{\partial t} + (\bfv \cdot \bm{\nabla})\bfv = -\frac{\bm{\nabla} P}{\rho} +\nu\nabla^2 \bfv + 
\bm{\Gamma},
 \label{eq:N_S_equ}
\end{equation}
where $P$, $\rho$, $\nu$ and $\bm{\Gamma}$ are the pressure, density, kinematic viscosity and extra force respectively. 
The three components of equation (\ref{eq:N_S_equ}) are
\begin{eqnarray}
 0 = -\frac{1}{\rho}\frac{\partial P}{\partial X} + \Gamma_X,
 \label{eq:x_comp}
\end{eqnarray}
\begin{eqnarray}
\begin{split}
 0 = -\frac{1}{\rho}\frac{\partial P}{\partial Y} + \nu \nabla^2 V_Y +\Gamma_Y,
 \label{eq:y_comp}
 \end{split}
\end{eqnarray}
\begin{eqnarray}
 0 = -\frac{1}{\rho}\frac{\partial P}{\partial Z} + \Gamma_Z.
 \label{eq:x_comp}
\end{eqnarray}

Equation (\ref{eq:y_comp}) can be further simplified to  
\begin{equation}
\begin{split}
\nabla^2 V_Y = \frac{1}{\nu}(-\Gamma_Y+\frac{1}{\rho}\frac{\partial P}{\partial Y}) = \frac{\partial^2 
V_Y}{\partial X^2} \\ 
\Rightarrow V_Y = -\left(\frac{\Gamma_Y}{\nu}-\frac{1}{\nu\rho}\frac{\partial P}{\partial Y}\right)\frac{X^2}{2} + 
C_1X+C_2& \\
 = -K\frac{X^2}{2} + C_1X + C_2, 
\end{split}
\end{equation}
where 
\begin{equation}
K = \left(\frac{\Gamma_Y}{\nu}-\frac{1}{\nu\rho}\frac{\partial 
P}{\partial Y}\right),
\label{kG}
\end{equation}
is assumed to be constant.

The corresponding boundary conditions are given by
\begin{equation}
 V_Y = \mp U_0\ {\rm at}\  X=\pm L,
 \label{eq:bc_dim_ful}
\end{equation}
which imply that $C_1 = -U_0/L$ and $C_2 = KL^2/2$. The background flow, therefore, in the presence of extra force modifies and it 
becomes

\begin{equation}
 V_Y = \frac{K}{2}(L^2-X^2)-\frac{U_0X}{L},  
 \label{eq:cou-PPF}
\end{equation}
which is nothing but Couette-Poiseuille flow, when linear and nonlinear 
shears both are present.
By a simple rearrangement, it reduces to
\begin{equation}
 V_Y = \alpha \left(1-\mathcal{X}^2\right),  
 \label{eq:PPF_reduction}
\end{equation} 
where 
\begin{equation}
\begin{split}
 \mathcal{X}^2 = \frac{K}{2}\frac{\Upsilon^2}{\alpha},\ \Upsilon^2 = \left(X+\frac{U_0}{KL}\right)^2 \\ \ {\rm and} \ \alpha 
= \frac{U_0^2}{2KL^2} + \frac{L^2K}{2}.
\label{eq:varible_def}
\end{split}
\end{equation}

The velocity $V_Y$ in equation (\ref{eq:PPF_reduction}) can be made dimensionless by dividing it with $\alpha$, i.e.
\begin{equation}
 U_{\alpha Y} = \frac{V_Y}{\alpha} = 1-\mathcal{X}^2,
 \label{eq:dim_less_velo}
\end{equation}
where $\alpha$ is the dimension of velocity, determined by the box geometry.
The new background flow, therefore, becomes 
$\bfu_{\alpha} = (0,U_{\alpha Y},0)$.
However, this is nothing but plane Poiseuille flow in new coordinates $(\mathcal{X}, Y, Z)$, where the boundary conditions 
are given by equation (\ref{eq:bc_dim_ful}). Note that here $\mathcal{X}$ is dimensionless, while $Y$ and 
$Z$ are dimensionful coordinates. Nevertheless, it is useful to solve
the problem within the known domain of the Poiseuille flow, 
i.e. $\mathcal{X}\in [-1, 1]$, in which it is known to be unstable above
certain Reynolds number ($Re$). 

\section{The new domain}
\label{sec:The new domain}
In order to employ the results of well-known Poiseuille flow, we set the 
boundary conditions in the new coordinates: 
$\bfu_{\alpha} = 0$ at $\mathcal{X} = \pm1$. It is therefore important 
to verify, the consequence of the domain of $\mathcal{X}$ (i.e., running from
-1 to 1) to the domain of $X$ (i.e., running from $-L$ to $L$), as chosen originally. From equation 
(\ref{eq:varible_def}), imposing $\mathcal{X}=\pm1$, we have
\begin{equation}
 X = \pm\sqrt{\frac{2\alpha}{K}} - \frac{U_0}{KL},
\end{equation}
where 
\begin{equation}
 \sqrt{\frac{2\alpha}{K}} = L\sqrt{\left(1+\frac{U_0^2}{K^2L^4}\right)}.
 \label{eq:2alphaoverK}
\end{equation}
Now if $U_0^2/K^2L^4 \ll 1$, equation (\ref{eq:2alphaoverK}) shows that
\begin{equation}
 \sqrt{\frac{2\alpha}{K}} \cong L+\frac{U_0^2}{2K^2L^3},
\end{equation}
leading to
\begin{equation}
 X = \pm L -\frac{U_0}{KL} \pm \frac{U_0^2}{2K^2L^3}.
\end{equation}
This confirms that the domain size of $X$ is close to $2L$ if $U_0^2/K^2L^4 \ll 1$, 
i.e. $\nu^2U_0^2/L^4\ll\Gamma_Y^2$, 
when the flow is not driven by the pressure, i.e. $\partial P/\partial Y = 0$. However, in the presence of pressure gradient,
the same condition will be true except its contribution will be added to the extra force.

However, if $U_0^2/K^2L^4 \gg 1$, equation (\ref{eq:2alphaoverK}) shows that
\begin{equation}
 \sqrt{\frac{2\alpha}{K}} \cong \frac{U_0}{KL}.
\end{equation}
Hence
\begin{equation}
 X = \pm\frac{U_0}{KL} - \frac{U_0}{KL},
\end{equation}
i.e. the domain size of $X$ is approximately $2U_0/KL$. According to the approximation $U_0^2/K^2L^4 \gg 1$, $2U_0/KL$ is
much larger than $2L$. Hence, domain increases compared to what chosen originally. Therefore, our original choice 
of a small Cartesian patch in the flow may violate with this choice. This further may create problem for 
application to accretion disks, described below in detail.
 
\begin{figure}
\includegraphics[width=\columnwidth]{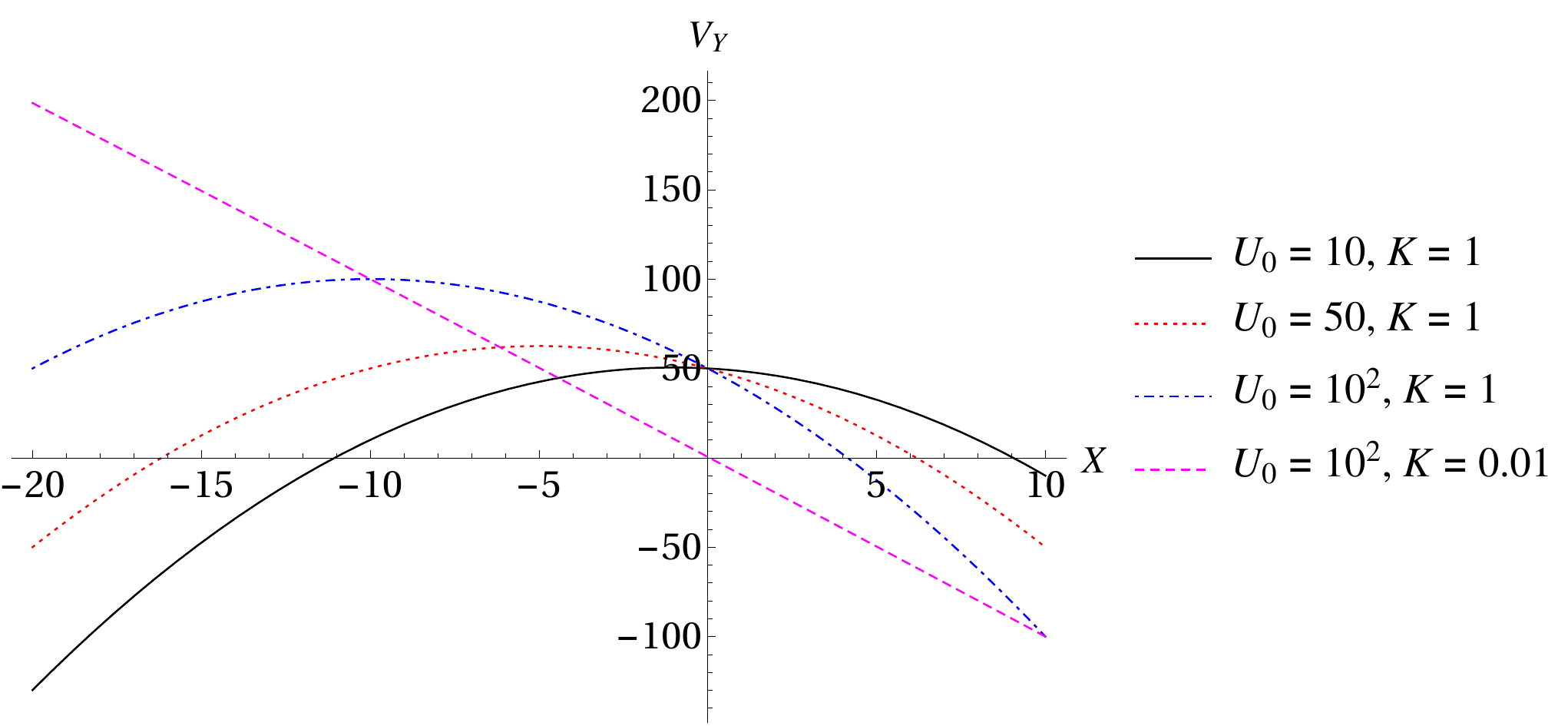}	  
  \caption{Modification of the background flow in the presence of a constant extra force is shown against the flow 
variable 
$X$ for various cases of parameters $U_0$ and $K$, when $L = 10.$}
\label{fig:mod_base_flow}
\end{figure}

FIG.~\ref{fig:mod_base_flow} shows the modified background flow in the presence of constant extra force for various $U_0$ 
and $K$ with $L = 10$. It is clear that for $U_0 = 10$ and $K=1$, the new domain
size almost remains same as $2L = 20$, because $U_0^2/K^2L^4 = 10^{-2} \ll 1.$ However, when $U_0$ increases keeping $K$
fixed, the domain size also increases. All these cases show the parabolic background flows in which the focus changes from
$(0,0)$ to $(-U_0/KL,0)$. Interestingly, if we keep decreasing $K$ keeping $U_0$ fixed, the linear background 
velocity eventually emerges again, as the extra force is negligible in this case (see equation \ref{eq:PPF_reduction}). 

The approximation $U_0^2/K^2L^4 \ll 1$ is, therefore, more suitable for our problem as it leads to almost the same domain size, 
$2L$, as chosen originally.

\section{Navier-Stokes equation in rotating frame}
\label{sec:N-S_eq_rot_frame}
Now our primary interest is to understand stability of
rotating shear flows, particularly in the context of accretion disks.
Hence, the plan is to examine the stability of the background flow with velocity $\bfu_{\alpha}$ within the new domain, i.e. 
$\mathcal{X}\in [-1,1]$, as discussed in 
\S\ref{sec:The new domain}, which is rotating w.r.to the center of the system 
at $R_0$.
Before doing so, we have to establish suitable 
equations depending on the reference
frame and make them dimensionless for the convenience. This exploration is essentially the stability
analysis of plane Poiseuille flow in the presence of rotation, here particularly the Coriolis effect. 
In the due process, 
we obtain a dimensionless number characterizing the flow to 
be laminar or turbulent in the domain of interest, i.e. the 
Reynolds number in the new coordinate system $(\mathcal{X},Y,Z)$ which is $Re_{\alpha}$.

\subsection{Defining Reynolds number in a local region}
\label{sec:Defining new Reynolds number}

Let us consider the local Cartesian frame or box at radius $R_0$, as shown by FIG. \ref{fig:shearing_box_pic}, 
rotating with the 
angular velocity $\bm{\omega} = (0,0,\Omega_0)$ such that $\Omega=\Omega_0(R/R_0)^{-q}$, where 
$\Omega_0 = V_{Y}(\mathcal{X} = 0)/(qL_{ch}/2) =  2\alpha\sqrt{K}/q\sqrt{2\alpha} = 
\alpha\sqrt{2K}/q\sqrt{\alpha}$, $L_{ch}$ is the characteristic length scale
of the system. 
The Navier-Stokes equation in that frame is
\begin{equation}
 \frac{\partial \bfv}{\partial t'} + (\bfv \cdot \bm\nabla)\bfv + \bm{\omega}\times\bm{\omega}\times \textbf{D} 
+2\bm{\omega}
 \times\bfv +\frac{\bm\nabla P}{\rho} = \nu \nabla^2\bfv+\bm{\Gamma},
\label{eq:NS_rot}
\end{equation}
where the position vector $\textbf{D} = (X,Y,Z)$, and $\bm\nabla = (\partial/\partial X, 
\partial/\partial Y, \partial/\partial Z)$ in Cartesian coordinates. If we divide both sides of equation 
(\ref{eq:NS_rot}) by $\alpha$, we obtain an equation for $\bfu_{\alpha}$, which is given by
\begin{equation}
\begin{split}
 \frac{\partial \bfu_{\alpha}}{\partial t'} + \alpha(\bfu_{\alpha} \cdot \bm\nabla)\bfu_{\alpha} 
 + \bm{\omega}\times\bm{\omega}\times \frac{\textbf{D}}{\alpha} +2\bm{\omega}\times\bfu_{\alpha}\\+\bm\nabla P_{\alpha} = 
 \nu \nabla^2\bfu_{\alpha}+\bm{\Gamma}_\alpha,
\label{eq:U_alpha_rot}
\end{split}
\end{equation}
where $P_{\alpha} = P/\rho \alpha$ and $\bm{\Gamma}_\alpha=\bm{\Gamma}/\alpha$. We now redefine the variables in terms of
dimensionless quantities, i.e. $t' 
\rightarrow 
\sqrt{2/\alpha K} t$, $(X,Y,Z) \rightarrow \sqrt{2\alpha/K} (\mathcal{X}, y, z)$, where $\textbf{d}=(\mathcal{X}, y, z)$, 
and $\bm\nabla \rightarrow \sqrt{K/2\alpha}\bm\nabla_{\alpha}$, where 
$\bm\nabla_{\alpha}=(\partial/\partial \mathcal{X}, \partial/\partial y, \partial/\partial z)$. Hence, equation 
(\ref{eq:U_alpha_rot}) in terms of 
dimensionless variables, for incompressible fluid, becomes
\begin{equation}
\begin{split}
 \frac{\partial \bfu_{\alpha}}{\partial t} + (\bfu_{\alpha} \cdot \bm\nabla_{\alpha})\bfu_{\alpha} +\frac{1}{q^2} 
\hat{k}\times\hat{k}\times\textbf{d} +\frac{2}{q}\hat{k}\times\bfu_{\alpha}\\+\frac{1}{\alpha}\bm\nabla_{\alpha} 
P_{\alpha} = \frac{\nu\sqrt{K}}{\alpha\sqrt{2\alpha}} 
\nabla_{\alpha}^2\bfu_{\alpha}+\Gamma_\alpha^\prime.
\label{eq:dim_less_NS}
\end{split}
\end{equation}
The Reynolds number, therefore, is defined as 
\begin{equation}
 Re_{\alpha} = \frac{\alpha\sqrt{2\alpha}}{\nu\sqrt{K}}
 \label{eq:Re_alpha}
\end{equation}
and $\bm{\Gamma}_\alpha^\prime=\bm{\Gamma}\sqrt{2/K\alpha^3}$.

\subsection{Perturbation analysis}
\label{sec:Perturbation analysis}
Equation (\ref{eq:dim_less_NS}) along with 
\begin{equation}
 \bm\nabla_{\alpha} \cdot \bfu_{\alpha} = 0
 \label{eq:div_U_alpha}
\end{equation}
describes the dynamics of fluid inside the local box. Now we perturb equations (\ref{eq:dim_less_NS}) and 
(\ref{eq:div_U_alpha}) linearly and check whether the perturbation decays or grows with time. The velocity perturbation 
is 
$\textbf{u}' = (u,v,w)$, the corresponding vorticity perturbation is $\bm\nabla \times \textbf{u}'$ and the pressure 
perturbation is $p'$. After perturbing equation (\ref{eq:dim_less_NS}), we eliminate the pressure term from the governing 
equation and recast it into the corresponding homogeneous Orr-Sommerfeld and Squire equations, which are 
given by  
\begin{eqnarray}
 \begin{split}
  \left(\frac{\partial}{\partial t} + U_{\alpha Y}\frac{\partial}{\partial y} \right)\nabla_{\alpha}^2u 
- U_{\alpha Y}''\frac{\partial u}{\partial y}+\frac{2}{q}\frac{\partial \zeta}{\partial z} \\- 
\frac{1}{Re_{\alpha}}\nabla_{\alpha}^4u  
= 0,\\
\label{eq:Orr_sommerfeld}
 \end{split}
\end{eqnarray}

\begin{eqnarray}
 \begin{split}
\left(\frac{\partial}{\partial t} + U_{\alpha Y}\frac{\partial}{\partial y} \right)\zeta - U_{\alpha Y}'\frac{\partial 
u}{\partial z}-\frac{2}{q}\frac{\partial u}{\partial z} \\- \frac{1}{Re_{\alpha}}\nabla_{\alpha}^2\zeta 
= 0,
\label{eq:Squire}
 \end{split}
\end{eqnarray}
where $\zeta$ is the $x$-components of the vorticity perturbations, prime denotes the differentiation 
w.r.to $\mathcal{X}$ and the extra force $\Gamma$ is assumed to 
remain same under perturbation so that it gets eliminated from the equation. 
If $\Gamma$ would have considered to be changed under 
perturbation, it would create additional
impact into the flow in order to reveal instability, as discussed by us recently (\citealt{our_work}) in the context
of linear shear. However, for the present purpose, we plan to explore minimum impact of force onto the flow.
Nevertheless, equations (\ref{eq:Orr_sommerfeld}) and (\ref{eq:Squire}) are the homogeneous part of the 
Orr-Sommerfeld and Squire equations corresponding to the perturbation to 
the equation (\ref{eq:dim_less_NS}) (see \citealt{our_work}, for inhomogeneous
Orr-Sommerfeld and Squire equations due to the effect of force). The corresponding chosen no-slip boundary conditions are: $u = v = w = 0$ at the two boundaries $\mathcal{X}=\pm1$, or equivalently $u = \frac{\partial u}{\partial 
\mathcal{X}} = \zeta = 0$ at $\mathcal{X}=\pm1$. Hence, the linearized eigenspectra 
corresponding to equation (\ref{eq:dim_less_NS}) will be described by 
equations (\ref{eq:Orr_sommerfeld}) and (\ref{eq:Squire}) only. The eigenspectra 
will not change due to the presence of the nonhomogeneous term arising due to the presence of the extra force. Our main 
aim here is to observe the changes in the eigenspectra, because of the changes in various flow properties.
Note importantly that in principle a small section of an accretion disk should not
have any boundary, as imposed here in order to introduce the boundary condition
for the solution purpose. However, the idea is that the entire disk, at least the region
where turbulence is sought of, is divided into small boxes and if one box is unstable
under perturbation, others will do so. All boxes are assumed to be arranged together.
Hence, on either side of a boundary, the perturbation remains working intact in the
respective boxes. Therefore, although boundaries are introduced for the solution
purpose, it does not practically introduce any artifact for the present purpose.
Nevertheless, \cite{man_2005} and \cite{amn_2005} showed that
results practically do not depend on whether the analyses are based on the shearing
sheet or shearing box.

\section{Perturbation analysis of rotating Poiseuille flow}
\label{sec:per-poi}
\subsection{Threedimensional perturbation}
\label{sec:Three dimensional perturbation}
In order to understand the evolution of linear perturbation,
let the linear solutions be (e.g. \citealt{man_2005})
\begin{eqnarray}
 u = \hat u(\mathcal{X},t) e^{i\textbf{k}\cdot \textbf{r}},
 \label{eq:trail_u}
 \\ 
 \zeta = \hat{\zeta}(\mathcal{X},t) e^{i\textbf{k}\cdot \textbf{r}},
  \label{eq:trail_zeta}
\end{eqnarray}
with $\textbf{k} = (0, k_y, k_z)\ {\rm and}\ \textbf{r} = (0, y, z).$
Substituting these in equations (\ref{eq:Orr_sommerfeld}) and (\ref{eq:Squire}), neglecting non-linear terms, we 
obtain
\begin{eqnarray}
 \begin{split}
  \frac{\partial \hat {u}}{\partial t} + i(\mathcal{D}^2 - k^2)^{-1}\Bigl[k_yU_{\alpha Y}\left(\mathcal{D}^2-k^2\right) - 
k_yU_{\alpha Y}''\\-\frac{1}{iRe_{\alpha}}(\mathcal{D}^2 - k^2)^2\Bigr]\hat{u} +(\mathcal{D}^2 - 
k^2)^{-1}\frac{2}{q}ik_z\hat{\zeta} = 0.
\label{eq:linear_orr_sommerfeld_eq_recast}
 \end{split}
\end{eqnarray}
and
\begin{equation}
 \begin{split}
  \frac{\partial \hat {\zeta}}{\partial t} 
+ik_yU_{\alpha Y}\hat{\zeta}-\left(U_{\alpha Y}'+\frac{2}{q}\right)ik_z\hat{u}-\frac{1}{Re_{\alpha}}(\mathcal{D}^2 - 
k^2)\hat{\zeta} =0,
\label{eq:linear_squire_eq}
 \end{split}
\end{equation}
where $\mathcal{D} = \frac{\partial}{\partial \mathcal{X}}$.

Further combining equations (\ref{eq:linear_orr_sommerfeld_eq_recast}) and (\ref{eq:linear_squire_eq}) we obtain 
\begin{eqnarray}
 \begin{split}
  \frac{\partial}{\partial t}Q +i\mathcal{L}Q = 0,
  \label{eq:lin_Q_eq}
 \end{split}
\end{eqnarray}
where
\begin{eqnarray}
\begin{split}
 Q = \begin{pmatrix}\hat{u}\\ \hat{\zeta}\end{pmatrix}, \ 
 \mathcal{L} = \begin{pmatrix}
                \mathcal{L}_{11}& \mathcal{L}_{12}\\
                \mathcal{L}_{21}& \mathcal{L}_{22}
               \end{pmatrix},
\label{eq:the_L_matrix}
\end{split}
\end{eqnarray}

\begin{eqnarray}
\begin{split}
 \mathcal{L}_{11} =& (\mathcal{D}^2 - 
k^2)^{-1}\Bigl[k_yU_{\alpha 
Y}\left(\mathcal{D}^2-k^2\right)-k_yU_{\alpha Y}''\\&-\frac{1}{iRe_{\alpha}}\left(\mathcal{D}^2-k^2\right)^2\Bigr],\\
 \mathcal{L}_{12} =& \frac{2k_z}{q}\left(\mathcal{D}^2-k^2\right)^{-1},\\
 \mathcal{L}_{21} =& -\left(U_{\alpha Y}'+\frac{2}{q}\right)k_z,\\
 \mathcal{L}_{22} =& k_yU_{\alpha Y}-\frac{1}{iRe_{\alpha}}\left(\mathcal{D}^2-k^2\right),
 \label{eq:L_diff_elements}
\end{split}
\end{eqnarray}

Now the solution of the equation (\ref{eq:lin_Q_eq}) is given by 
\begin{eqnarray}
 Q = \sum_{m = 1}^{\infty}C_m Q_{\mathcal{X},m}(\mathcal{X}) \exp{(-i\sigma_m t)},
 \label{eq:gen_sol_Q}
\end{eqnarray}
where index $m$ corresponds to the combined Orr-Sommerfeld and Squire modes and  $Q_{\mathcal{X},m}(x)$ 
satisfies the eigenvalue 
equation
\begin{eqnarray}
 \mathcal{L}Q_{\mathcal{X},m}(\mathcal{X}) = \sigma_m 
Q_{\mathcal{X},m}.
 \label{eq:eival_eq}
\end{eqnarray}
Here $\sigma$ is a complex quantity, given by $\sigma = \sigma_r + i \sigma_i$. In this article,
$\sigma_r$, $\sigma_i$ and hence $\sigma$ for different parameters and different flows are obtained numerically. 
The numerical method for discretization is described in \S\ref{sec:Numerical technique}.

\begin{figure}
\includegraphics[width=\columnwidth]{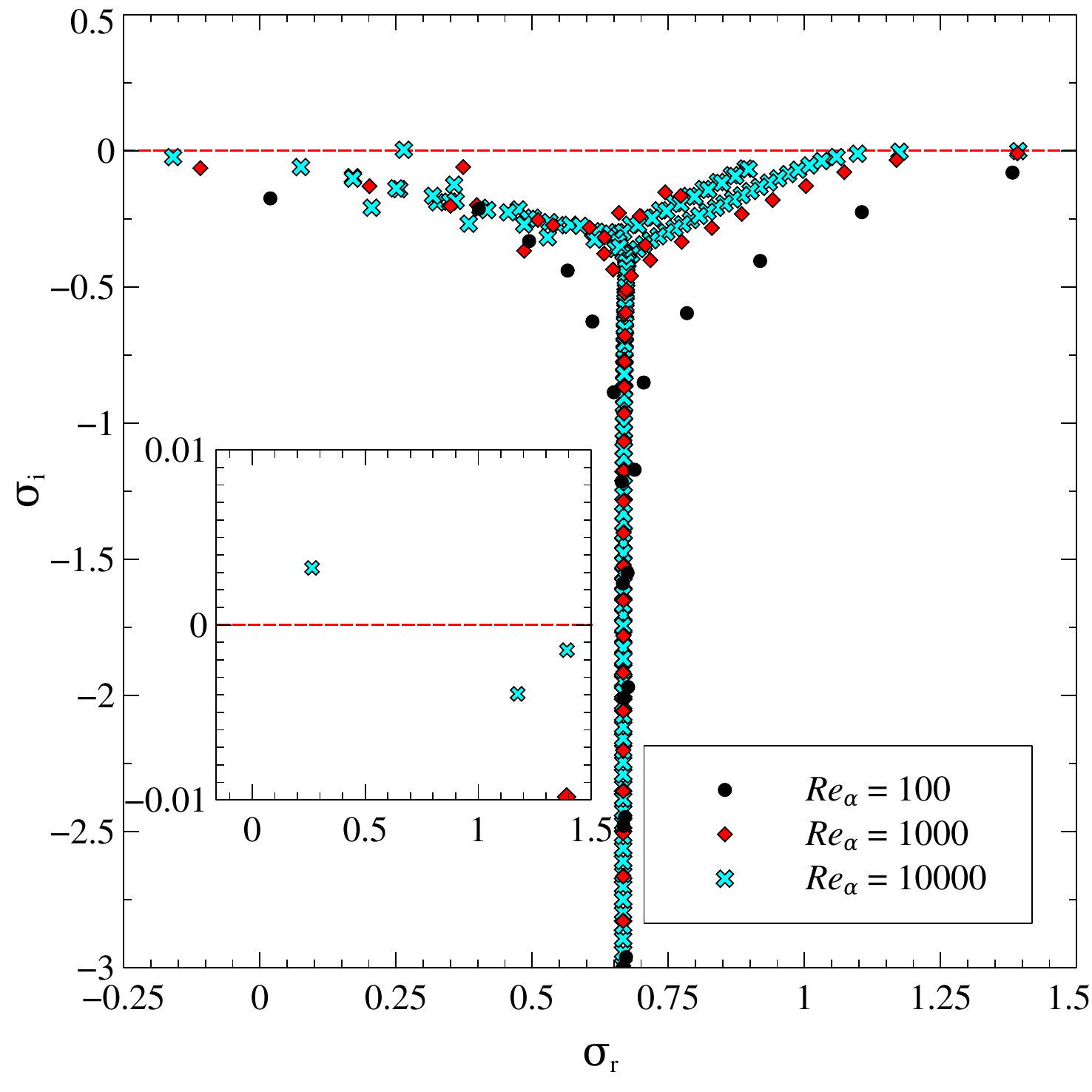}	 
  \caption{Eigenspectra for linearized Poiseuille flow in the presence of 
Keplerian rotation ($q=1.5$) of the box for $Re_{\alpha} = 100,\ 1000$ and 
$10000$, with $k_y = k_z = 1$.}
\label{fig:eval_poi_rot_three_diff_R_3d}
\end{figure}

\begin{figure}
\includegraphics[width=\columnwidth]{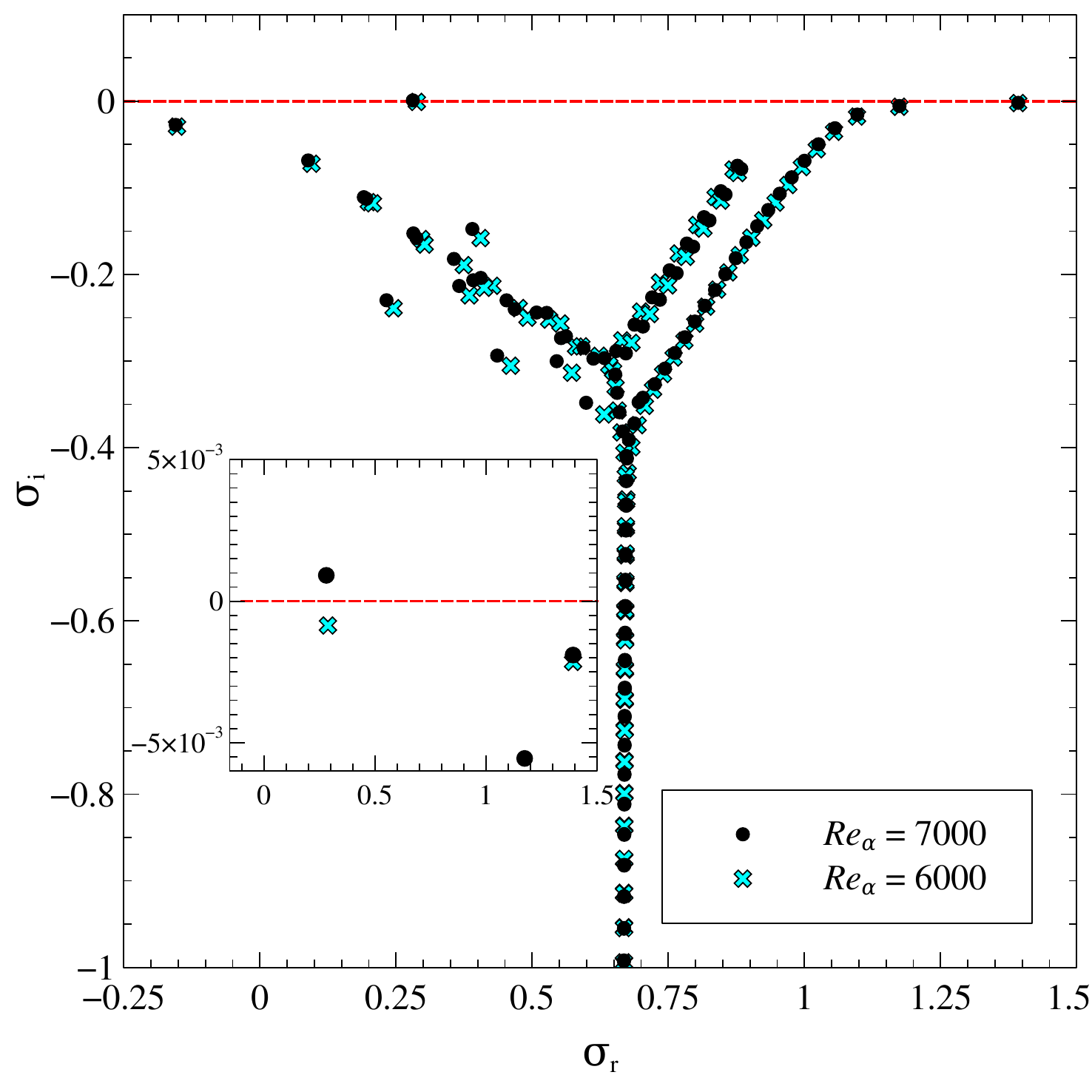}      
  \caption{Same as \ref{fig:eval_poi_rot_three_diff_R_3d}, except
for $Re_{\alpha} = 6000$ and $7000$.}
\label{fig:eval_poi_rot_diff_R_3d}
\end{figure}

\begin{figure}
	\includegraphics[width=\columnwidth]{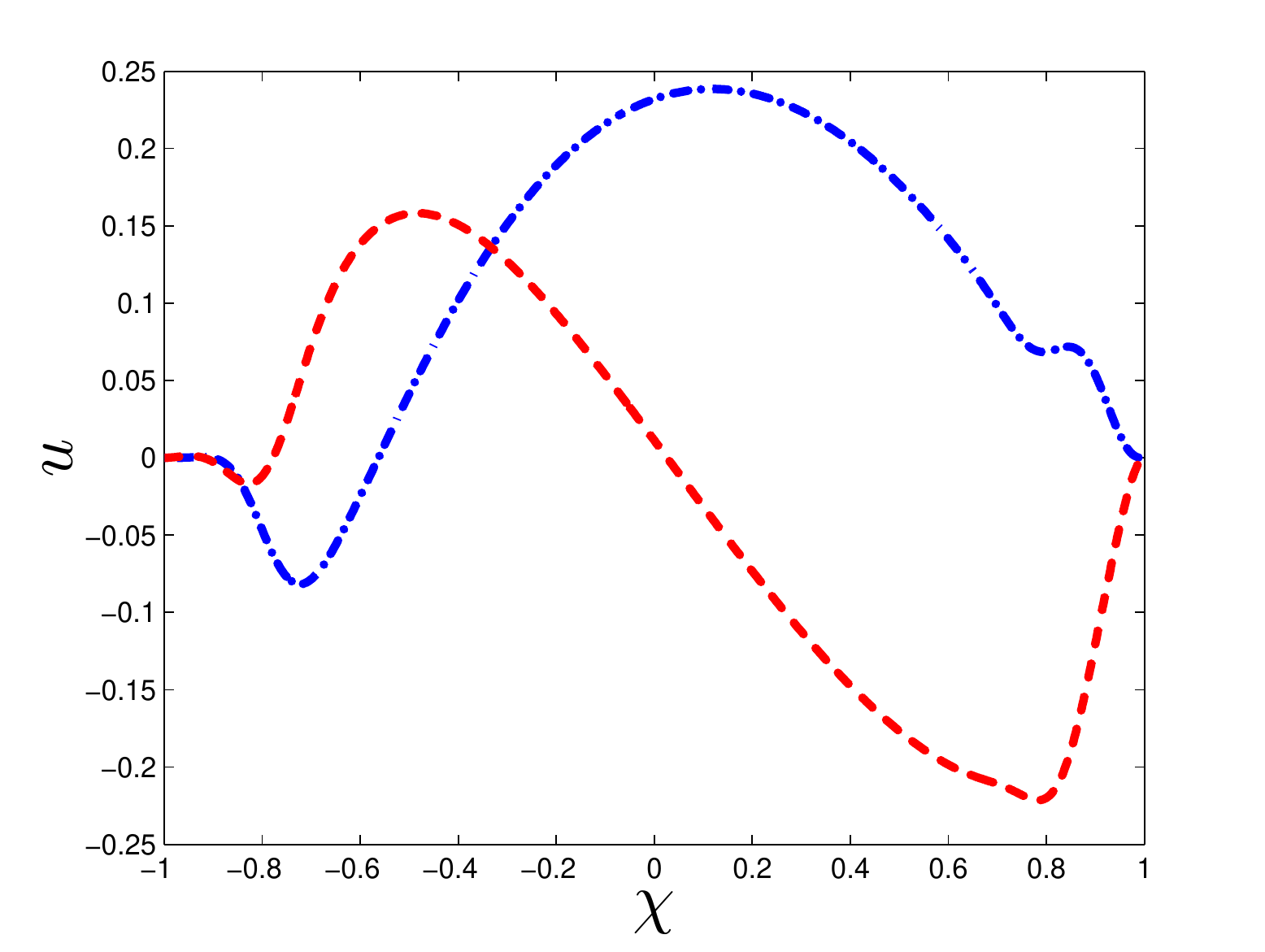}	 
  \caption{Velocity eigenfunction for the most unstable mode 
	corresponding to linearized Poiseuille flow 
in the presence of Keplerian rotation ($q=1.5$) of the box 
	for $Re_{\alpha} = 7000$ with $k_y = k_z=1$. 
Dot-dashed and dashed lines indicate, respectively, the real and imaginary parts of $u$.}
\label{fig:evec_poi_rot_3d_re_7000_kep_velo}
\end{figure}

\begin{figure}
\includegraphics[width=\columnwidth]{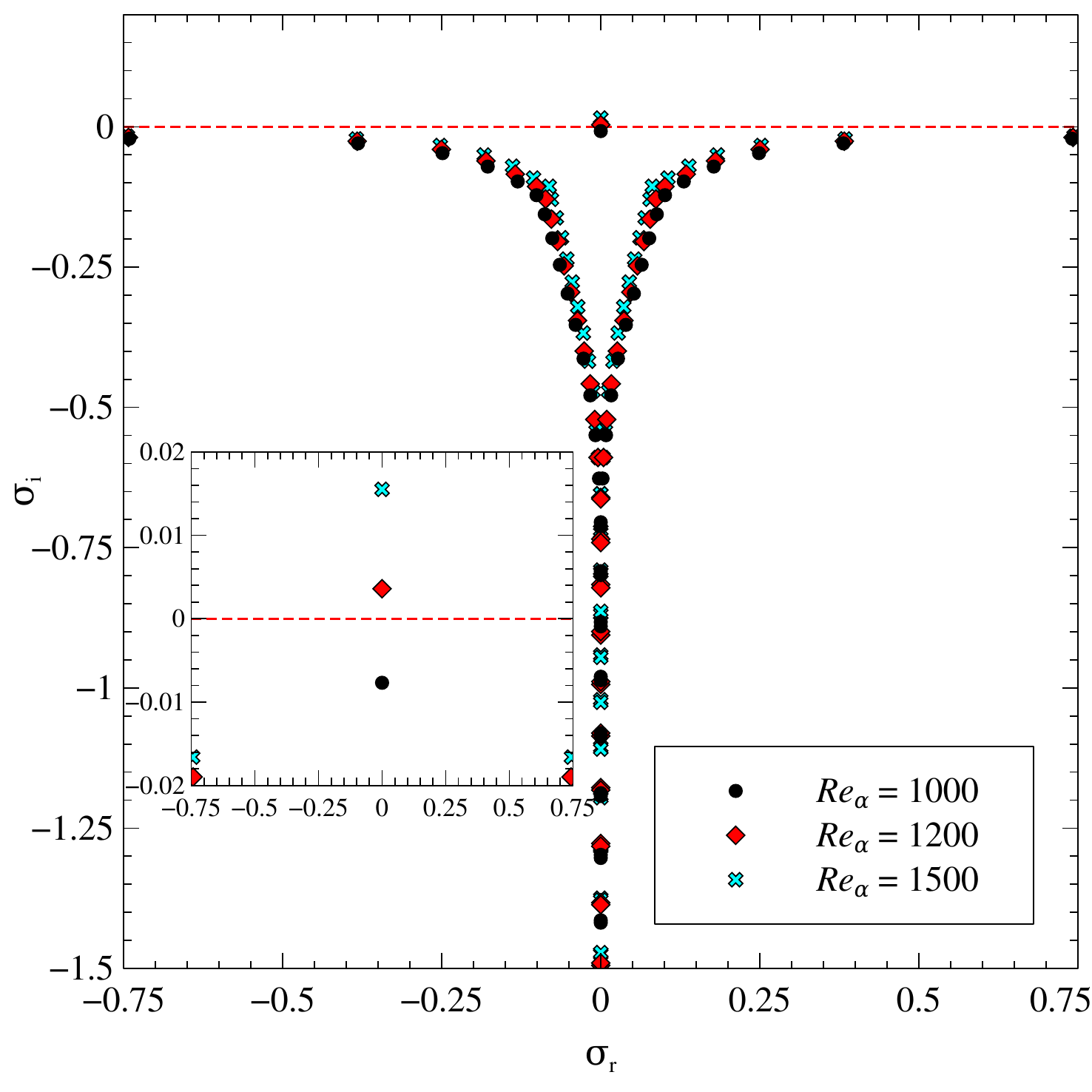}   
  \caption{Eigenspectra for linearized Poiseuille flow in the presence of 
Keplerian rotation ($q=1.5$) of the box for $Re_{\alpha} = 1000,\ 1200$ and 
$1500$, with a
pure vertical perturbation, i.e. $k_y = 0$ and $k_z = 1$.}
\label{fig:eval_poi_rot_diff_R_3d_pure}
\end{figure}

It is well-known that Poiseuille flow is linearly unstable under twodimensional 
perturbation with $k_y=1.02$ and $k_z=0$ and the critical $Re$ 
is about 5772.22. However for the same $Re$, this flow is stable 
under perturbation with $k_y=k_z=1$. Nevertheless, in the presence of Keplerian rotation of the whole system,
i.e. due to the effect of Coriolis force in the local rotating box, the
situation changes.
FIG.~\ref{fig:eval_poi_rot_three_diff_R_3d} shows the eigenspectra for 
linearly perturbed Poiseuille flow in the rotating frame (while the small region under consideration is rotating) with 
$q = 1.5$ (Keplerian rotation)
for $k_y=k_z = 1$ for different $Re_{\alpha}$. Here we observe that 
while the flow is stable for 
$Re_{\alpha} = 100\ {\rm and} \ 1000$ with $k_y = k_z =1$, it is unstable 
for $Re_{\alpha} = 10000$. FIG.~\ref{fig:eval_poi_rot_diff_R_3d} also depicts 
the eigenspectra for the same flow as in FIG.~\ref{fig:eval_poi_rot_three_diff_R_3d} but for $Re_{\alpha} = 6000$ and 
$7000$. It confirms that instability arises between for $Re_{\alpha} = 6000$ and $7000$. 
The critical $Re_{\alpha}$ is around 6431.473. FIG.~\ref{fig:evec_poi_rot_3d_re_7000_kep_velo} 
depicts a sample of velocity eigenfunction, which is given here for the most 
unstable mode corresponding to Poiseuille flow for $Re_{\alpha} = 7000$, 
$k_y = k_z = 1$, and $q =1.5$. 

Depending on the localization of eigenfunctions, the corresponding modes are named. If the 
eigenfunctions have their maxima around the center of the domain, the corresponding modes are called body modes. If the 
eigenfunctions are localized around the boundary, the corresponding modes are called wall modes. See \cite{Kersale2004} 
for details about these modes. From FIG.~\ref{fig:evec_poi_rot_3d_re_7000_kep_velo}, we see that the modes 
are body modes.

We have imposed no-slip boundary conditions to obtain all the eigenspectra 
corresponding to Poiseuille flow in the presence of rotation. However, for different boundary conditions, the unstable 
nature of the flow does not disappear from the system. \cite{Xiong_2020} 
argued that with the change of boundary conditions, only the critical 
Reynolds number and other parameters revealing instability change.

To have a qualitative sense of why plane Poiseuille flow in the presence 
of Keplerian rotation becomes unstable under threedimensional perturbation unlike the nonrotating case, 
below we investigate the effect of pure vertical perturbation.

\subsection{Pure vertical perturbation}
\label{sec:Pure vertical perturbation}
We consider pure vertical perturbation of the form $u, \zeta \sim u(t), \zeta(t) \exp(ik_z z)$ for the ease of 
analytical exploration and hence equations (\ref{eq:Orr_sommerfeld}) and (\ref{eq:Squire}) reduce to respectively
\begin{equation}
 \frac{\partial u}{\partial t}- \frac{2i}{qk_z}\zeta = -\frac{1}{Re_{\alpha}}k_z^2u
 \label{eq:u_kz}
\end{equation}
and 
\begin{equation}
 \frac{\partial \zeta}{\partial t} - ik_z(\frac{2}{q} + U'_{\alpha Y})u = -\frac{1}{Re_{\alpha}}k_z^2\zeta.
 \label{eq:zeta_kz}
\end{equation}
Note, however, that this is just for the sake of an approximate 
analytical exploration, as due to shear in the $x$-direction, the perturbation
cannot have this form, as we even did not choose it in our exploration of 
eigenspectrum analysis.
Combining equations (\ref{eq:u_kz}) and (\ref{eq:zeta_kz}), we obtain a second order temporal differential equation 
for $u(t)$, given by
\begin{equation}
 \frac{\partial^2 u}{\partial t^2} + \frac{2k_z^2}{Re_{\alpha}}\frac{\partial u}{\partial t} + 
\left(\frac{4}{q^2}+\frac{2U'_{\alpha Y}}{q}+\frac{k_z^4}{Re_{\alpha}^2}\right)u = 0.
\label{eq:2nd_der_u}
\end{equation}

Let us consider the solution of equation (\ref{eq:2nd_der_u}) be $u(t)\sim \exp(\sigma t)$, $\gamma = 2k_z^2/Re_{\alpha}$ 
and $\beta = 4/q^2+2U'_{\alpha Y}/q + k_z^4/Re_{\alpha}^2$. Hence from equation (\ref{eq:2nd_der_u}), we obtain a quadratic equation for $\sigma$, given by
\begin{equation}
 \sigma^2 + \gamma \sigma + \beta = 0,
 \label{eq:quadratic_eq}
\end{equation}
whose solution is 
\begin{equation}
 \sigma = -\frac{\gamma}{2} \pm \sqrt{-\frac{4}{q^2} - \frac{2U'_{\alpha Y}}{q}}.
 \label{eq:gen_disp_rel}
\end{equation}
If the background flow follows the equation (\ref{eq:dim_less_velo}), then equation (\ref{eq:gen_disp_rel}) becomes
\begin{equation}
 \sigma = -\frac{\gamma}{2} \pm 2\sqrt{\frac{\mathcal{X}}{q}-\frac{1}{q^2}}.
 \label{eq:plane_Poi_disp_rel}
\end{equation}
From the above equation, it is clear that whenever the quantity under the root is real
positive, one of the solutions for the vertical
perturbation may grow with time exponentially. This depends on whether the magnitude of the term involved with 
square root (second term)
is greater or less than the first term. As $\mathcal{X}$ varies between $-1$ and $+1$, it is very obvious that the 
vertical perturbation will grow for the system under consideration. However as the growing modes 
correspond to a real positive $\sigma_i$, there is a lower bound of $\mathcal{X}$ for the modes to be confined in the system.

FIG.~\ref{fig:eval_poi_rot_diff_R_3d_pure} shows, by full numerical solutions, the eigenspectra for the Poiseuille 
flow in the presence of Keplerian 
rotation in the case of pure vertical perturbation, i.e. $k_y = 0\ {\rm and}\ k_z = 1$, for $Re_{\alpha} = 1000,\ 1200\ 
{\rm and}\ 1500$. From the inset figure, it is very clear that the flow is stable for $Re_{\alpha} = 1000$. However, it 
is unstable for $Re_{\alpha} = 1200\ {\rm and}\ 1500$ and 
the critical $Re_{\alpha}$ is around 1129.18.

Above results argue that for an astrophysical accretion disk, when the flow
is necessarily threedimensional with rotation, a very small force makes
the system unstable as its $Re$ is huge (see, e.g., \citealt{Mukho_2013}).
Even if $\partial P/\partial Y$ vanishes, a finite 
$\Gamma_Y/\nu\equiv Re_{\alpha}\Gamma_Y$ would suffice for instability due to the emergence of 
small contribution of $x^2$ (Poiseuille) effect along 
with $x$ (Couette) effect in the background.
In fact at large $U_0$, when the quadratic term in $X$ in equation (\ref{eq:cou-PPF}) 
is small compared to the linear term in $X$, $Re_\alpha\rightarrow \nu U_0^3/2L^3\Gamma_Y^2$. 
Therefore, a very small force $\Gamma_Y$ along with a similar small $\nu$ would suffice 
a huge $Re_\alpha$, which might lead to linear instability and subsequent
nonlinearity and turbulence in accretion disks. 
In plane Couette flow in laboratory, when $\nu$ is not very small, still
a small force would lead to instability and turbulence at large $U_0$.
For an intermediate $U_0$, the instability is expected to arise at an 
intermediate $Re_\alpha$, as seen in experiments.
The above arguments remain intact if the force is solely due 
to the unavoidable pressure gradient, whether tiny or not, such that $K=1/(\nu\rho)\partial P/\partial Y$. 
We will discuss in detail 
the relative importance of external force and background velocity along with viscosity
in order to control flow stability in \S \ref{sec:cou-poi} below.

\subsection{Dependence of the stability on rotation profile}
\label{sec:Dependence of rotation}

\begin{figure}
\includegraphics[width=\columnwidth]{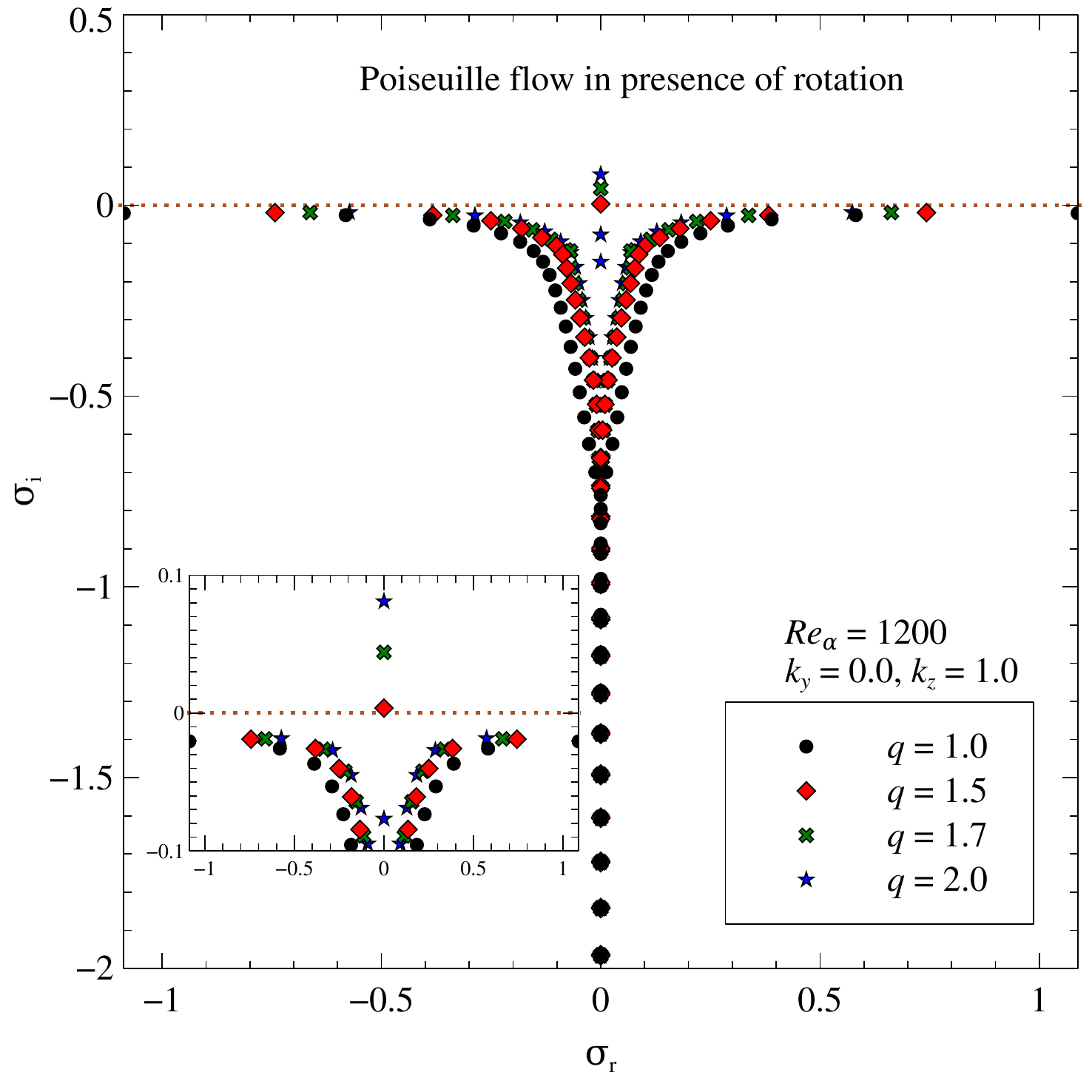}    
  \caption{Eigenspectra of linearized Poiseuille flow in the presence of rotation for vertical perturbation 
  with $k_y =0$ and $k_z = 1$ 
  for four different $q$ and $Re_{\alpha} = 1200$.}
\label{fig:Poi_rot_diff_q_3d_pure_Re_1200}
\end{figure}

\begin{figure}
\includegraphics[width=\columnwidth]{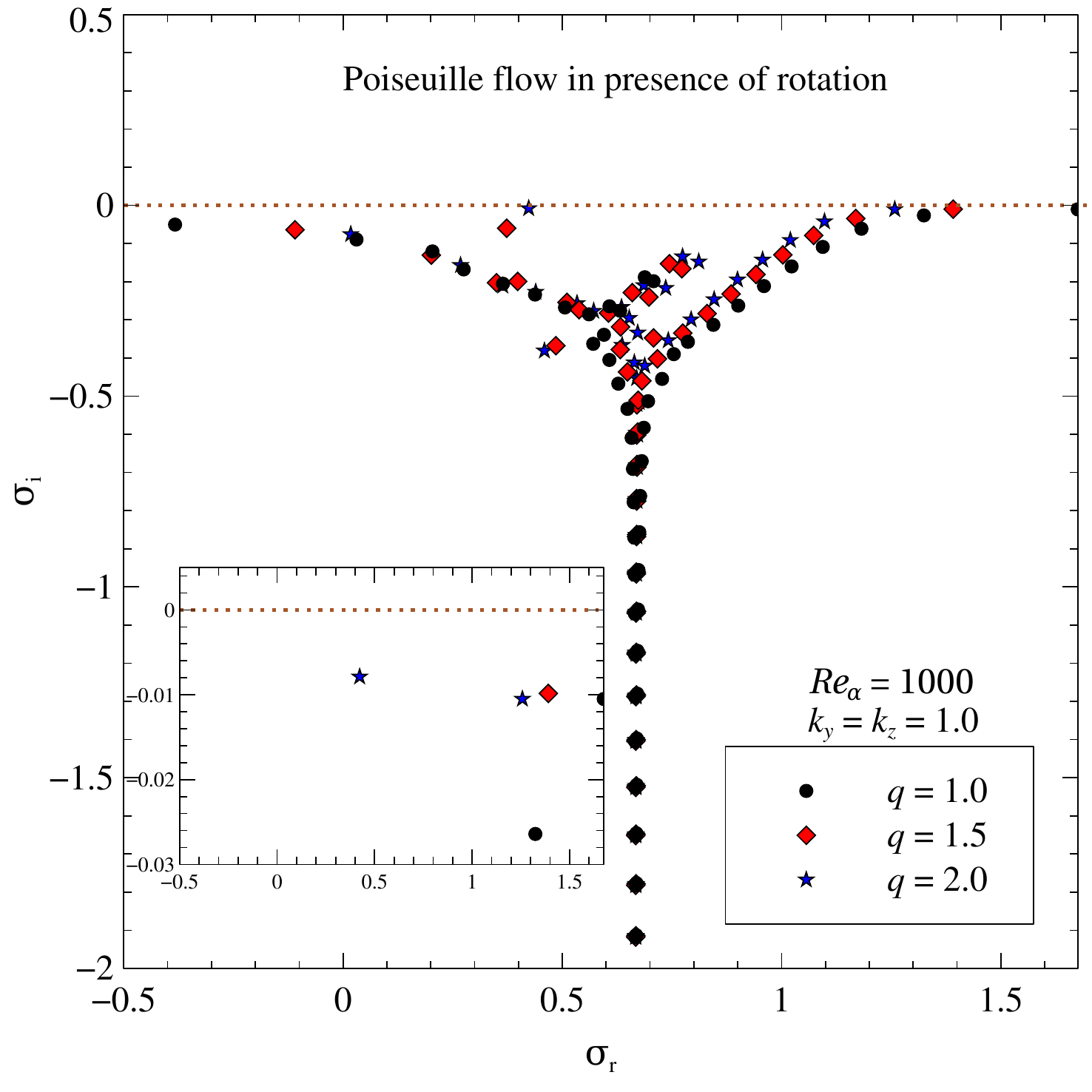}	
  \caption{Eigenspectra of linearized Poiseuille flow in the presence of rotation for threedimensional perturbation 
  with $k_y = k_z = 1$ for three different $q$ and $Re_{\alpha} = 1000$.}
\label{fig:Poi_rot_diff_q_3d_Re_1000}
\end{figure}

\begin{figure}
\includegraphics[width=\columnwidth]{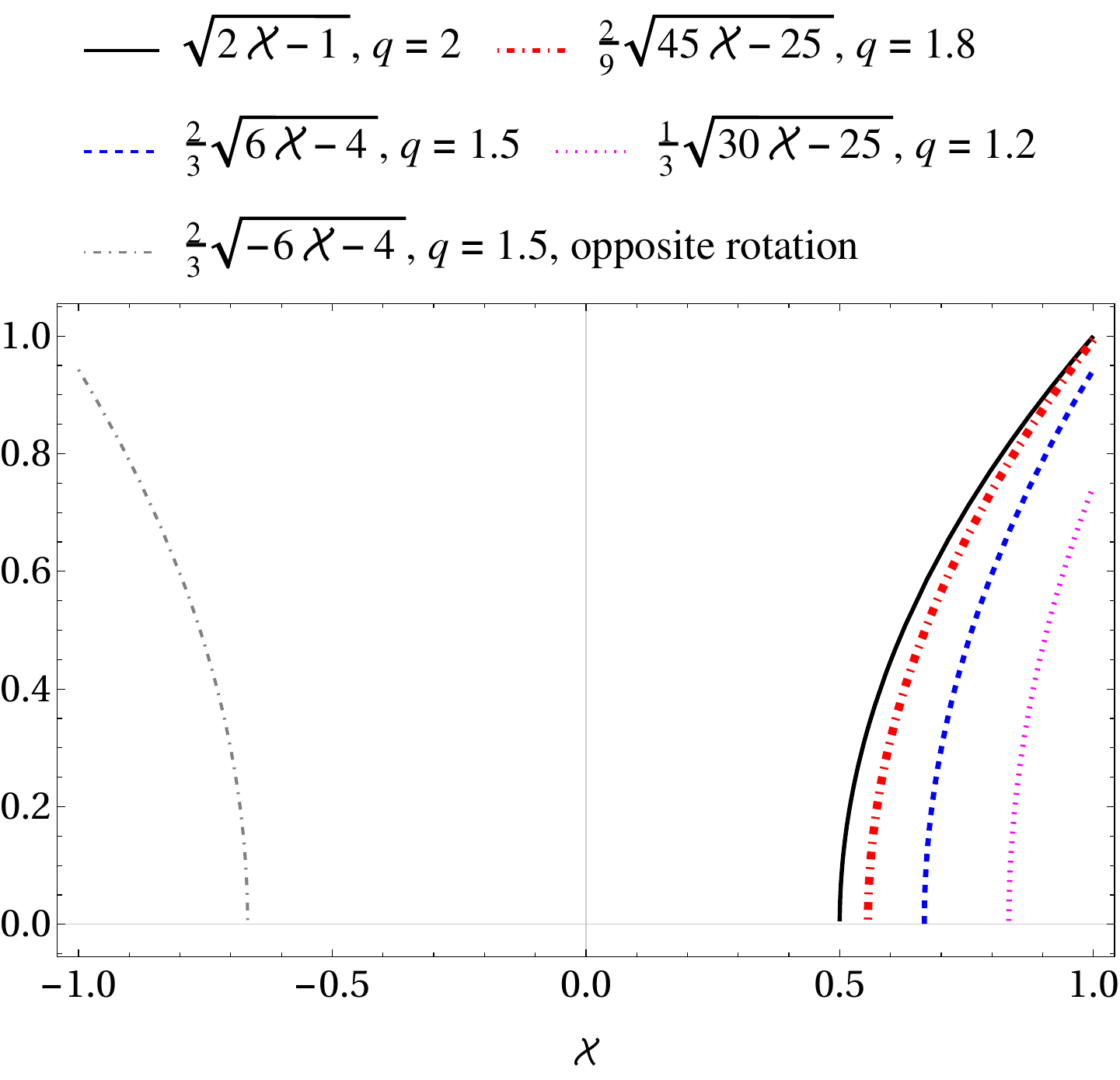}	
  \caption{Variation of $2\sqrt{\mathcal{X}/q-1/q^2}$ from equation (\ref{eq:plane_Poi_disp_rel}) as a function of
  $\mathcal{X}$ for $q = 1.2,\ 1.5,\ 1.8,\ 2.0$ and for a oppositely
rotating flow with $q=1.5$.}
\label{fig:stability}
\end{figure}

\begin{figure}
\includegraphics[width=\columnwidth]{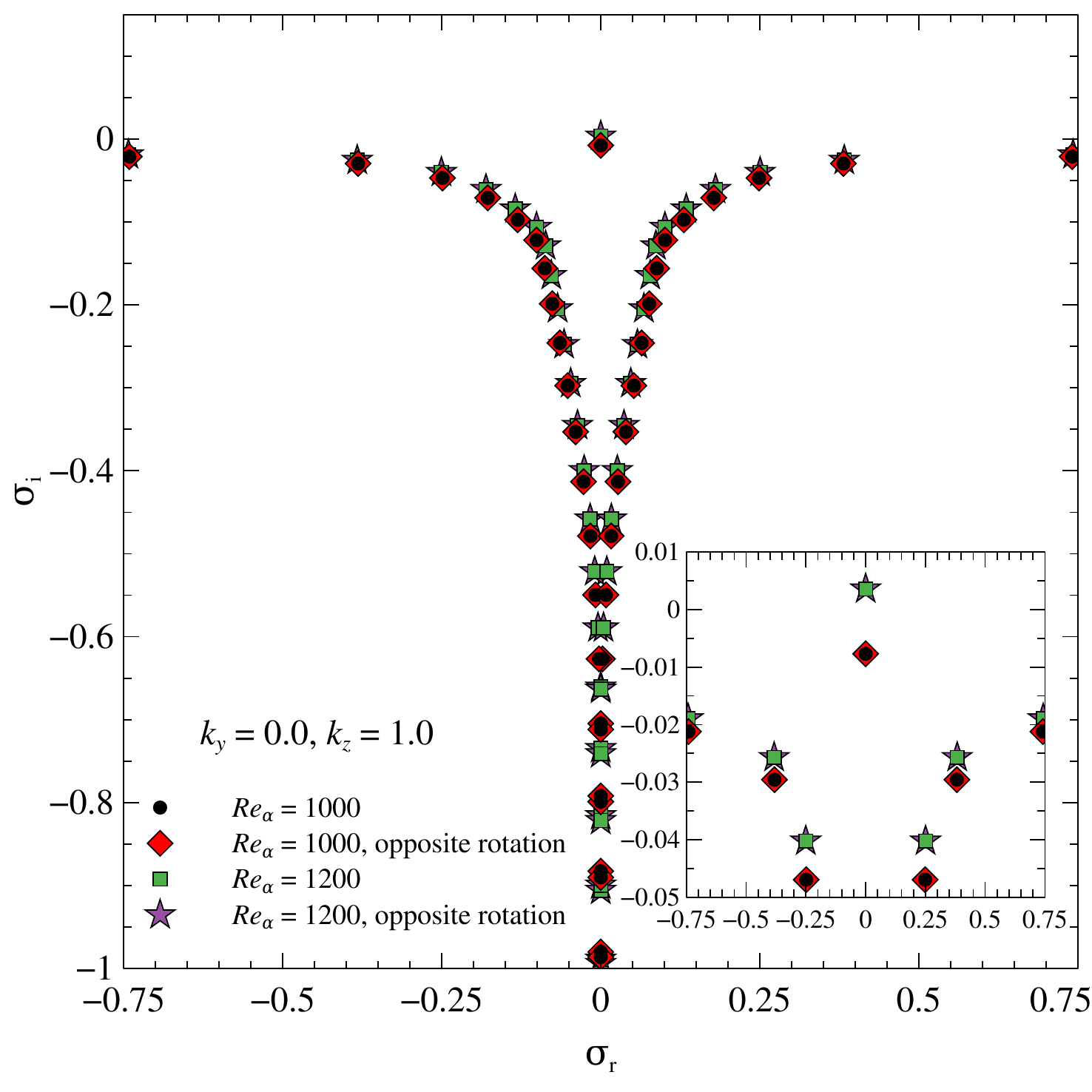}	
  \caption{Eigenspectra of linearized Poiseuille flow in the presence of rotation for vertical perturbation with $k_y = 0$, and 
  $k_z = 1$ for $Re_{\alpha}$, but for two opposite orientations of 
Keplerian rotation.}
\label{fig:Poi_rot_pos_neg_q_3d_pure_Re_1000_1200}
\end{figure}

\begin{figure}
\includegraphics[width=\columnwidth]{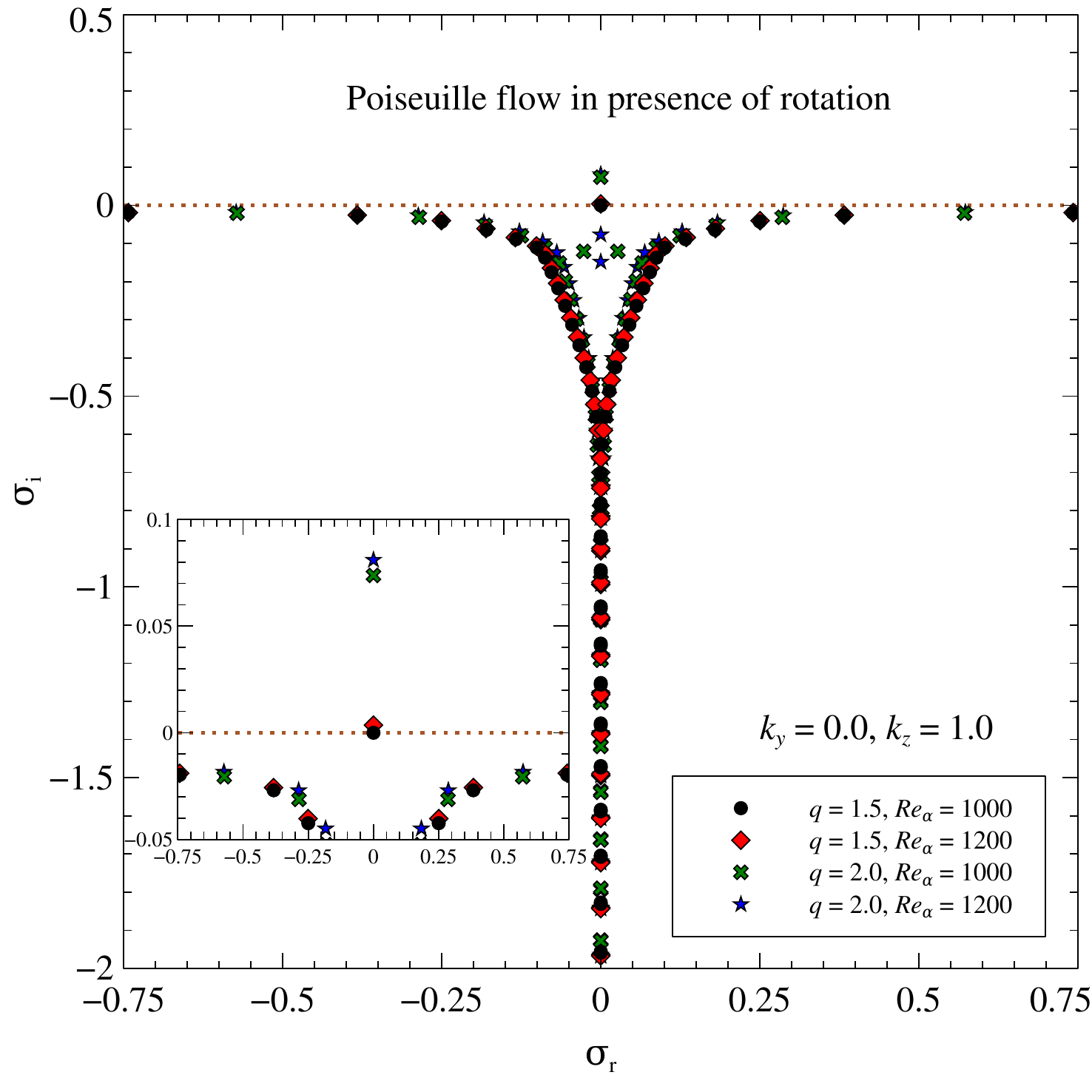}	
  \caption{Eigenspectra of linearized Poiseuille flow in the presence of rotation for vertical perturbation with $k_y = 0$, 
  and $k_z = 1$ for two different $q$ and $Re_{\alpha}$.}
\label{fig:Poi_rot_two_q_3d_pure_Re_1000_1200}
\end{figure}

\begin{figure}
\includegraphics[width=\columnwidth]{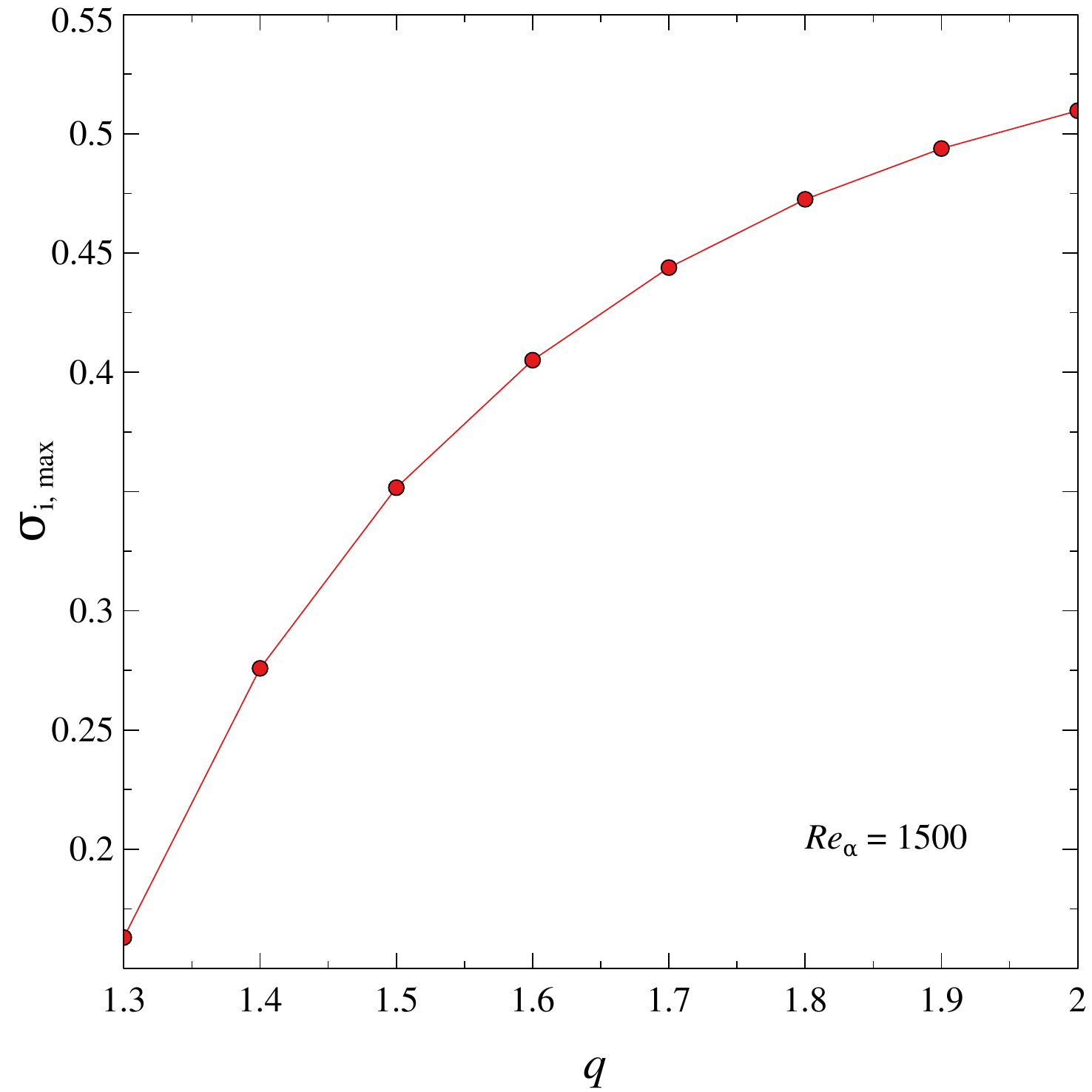}	
  \caption{Maximum growth rate ($\sigma_{i, max}$) as a function of $q$ in linearized Poiseuille flow
  for vertical perturbation with those $k_z$ giving rise to $\sigma_{i, max}$ for 
  $Re_{\alpha} = 1500$.}
\label{fig:poi_sigma_i_max_vs_q}
\end{figure}

\begin{figure}
\includegraphics[width=\columnwidth]{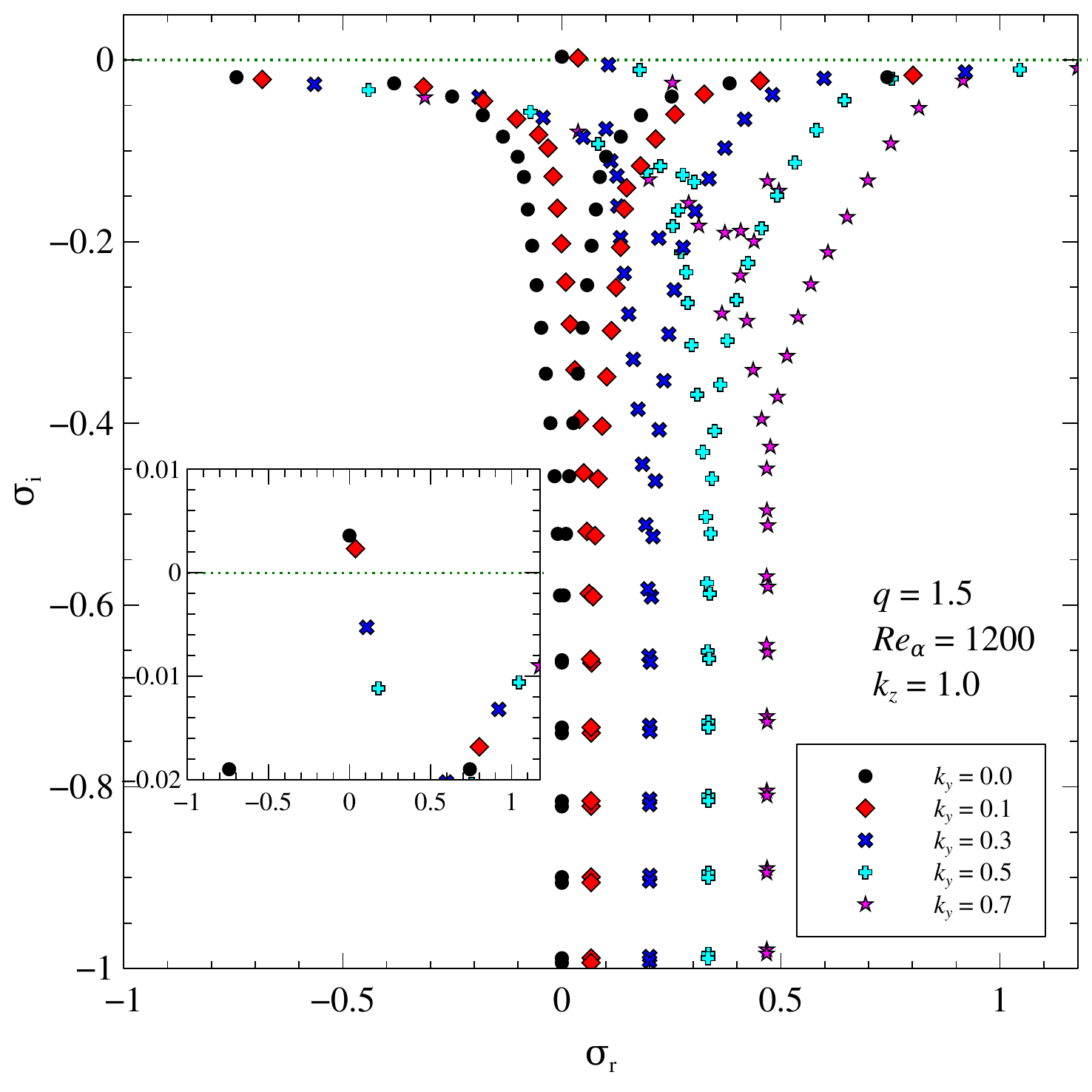}	
  \caption{Eigenspectra of linearized Poiseuille flow in the presence of Keplerian rotation for perturbations with $k_z = 1$ 
  but with 
  five different $k_y$ for $Re_{\alpha} = 1200$.}
\label{fig:Poi_rot_kep_3d_diff_ky_Re_1200}
\end{figure}

In the last two subsections, we have observed how the Keplerian rotation affects the threedimensional and vertical 
perturbations. Here, we shall observe  how the stability of plane Poiseuille flow depends on rotation profile, i.e. on 
different $q$. FIG.~\ref{fig:Poi_rot_diff_q_3d_pure_Re_1200} describes the eigenspectra of Poiseuille flow for three different rotational profiles
under a purely vertical perturbation. Here we notice that for a fixed 
$Re_{\alpha}$, larger $q$ has larger growth rate. In addition, we notice that while 
$q=1$ provides stable flow, the other three chosen $q$ result in unstable flow. For a three 
dimensional perturbation (i.e. $k_y$ and $k_z$ both nonzero) also stability
decreases with increasing $q$, as is evident from
FIG.~\ref{fig:Poi_rot_diff_q_3d_Re_1000}. 
However, as opposed to the purely vertical perturbation, in the threedimensional
case, flows of all $q$ are stable, for the chosen set of parameters. 

Let us understand this fact from equation (\ref{eq:plane_Poi_disp_rel}). We consider two extreme cases of $q$, i.e. $q=1$ 
and $q=2$. Equation (\ref{eq:plane_Poi_disp_rel}) for these two cases become
\begin{equation}
 \sigma = -\frac{\gamma}{2} \pm 2\sqrt{\mathcal{X} -1}
 \label{eq:plane_Poi_disp_rel_q_1}
\end{equation}
and 
\begin{equation}
 \sigma = -\frac{\gamma}{2} \pm \sqrt{2\mathcal{X} -1}
 \label{eq:plane_Poi_disp_rel_q_2}
\end{equation}
respectively. From equation (\ref{eq:plane_Poi_disp_rel_q_1}), we can obtain that the system will be unstable 
if $(\mathcal{X}-1)$ is a positive real number and 
\begin{equation}
 \sqrt{\mathcal{X} -1} > \frac{\gamma}{4} = \frac{k_z^2}{2Re_{\alpha}}.
 \label{eq:stability_cond_disp_q_1}
\end{equation}
Now, $(\mathcal{X} - 1)$ is negative other than at $\mathcal{X}=1$ as $\mathcal{X}\in [-1,1]$.
However at $\mathcal{X} = 1$, equation (\ref{eq:stability_cond_disp_q_1}) 
is also not valid there as $k^2/Re_{\alpha} > 0$ always. 
Therefore, plane Poiseuille flow with $q=1$ will
always be stable. It explains the reason
behind the stable Poiseuille flow with vertical perturbations for $q = 1$ as depicted in the 
FIG.~\ref{fig:Poi_rot_diff_q_3d_pure_Re_1200}. 

To make the system unstable for $q = 2$, from equation (\ref{eq:plane_Poi_disp_rel_q_2}), we require 
$(2\mathcal{X} -1)$ to be a positive real number and  
\begin{equation}
 \sqrt{2\mathcal{X} -1} > \frac{\gamma}{2} = \frac{k_z^2}{Re_{\alpha}}.
 \label{eq:stability_cond_disp_q_2}
\end{equation}
Now $(2\mathcal{X} -1)$ could be positive within the domain of $\mathcal{X}$. Hence, for
$q=2$, plane Poiseuille flow will be unstable depending on the parameters, i.e. $k_z$ and $Re_{\alpha}$.

Now for a general $q$ (usually $1\leq q \leq 2$ for the present interest), to make the system unstable, 
$\mathcal{X}/q - 1/q^2$ has to be a positive real number and also from equation 
(\ref{eq:plane_Poi_disp_rel})
\begin{equation}
 2\sqrt{\frac{\mathcal{X}}{q}-\frac{1}{q^2}}> \frac{\gamma}{2} = \frac{k_z^2}{Re_{\alpha}}.
 \label{eq:stability_cond_disp_gen}
\end{equation}
It is important to check in which domain of $\mathcal{X}$, $\mathcal{X}/q - 1/q^2$ is positive 
and equation (\ref{eq:stability_cond_disp_gen}) satisfies. Moreover, it is also important to know how the maximum 
growth rates for vertical perturbations depend on $q$. The answers to these queries can be found in FIG.~\ref{fig:stability},
where the variation of $2\sqrt{\mathcal{X}/q-1/q^2}$ as a function of $\mathcal{X}$ is shown for five different $q$. 
It shows that the size of the domain, in which $2\sqrt{\mathcal{X}/q-1/q^2}$ 
remains a real number and hence $(\mathcal{X}/q-1/q^2)$ remains a positive real number, decreases as $q$ decreases. 
From equation (\ref{eq:plane_Poi_disp_rel}), it is obvious that the maximum growth rates will be larger for those $q$'s for
which $2\sqrt{\mathcal{X}/q-1/q^2}$ will be larger. From FIG.~\ref{fig:stability}, we notice that as $q$ decrease, 
$2\sqrt{\mathcal{X}/q-1/q^2}$ also decreases. This reveals that 
larger $q$ will have larger growth rates with a larger domain. This 
explains the reason behind the larger growth rates (less stability) 
for vertical (threedimensional) perturbations for larger $q$ for a fixed $Re_{\alpha}$. 

Interestingly, above findings argue for a striking similarities between the Papaloizou-Pringle instability (PPI, see 
e.g. \citealt{Papaloizou_1984,Balbus_2003}) and the instability that we have obtained here, specifically the one with 
the vertical perturbations. For PPI, the pressure gradient is nonzero in 
the equilibrium. 
In our case, we also consider a modified background due to the presence of pressure gradient and/or an external force in the local 
frame. For PPI, the perturbations have to be non-axisymmetric. These 
non-axisymmetric perturbations contain the 
information of rotation (\citealt{Balbus_2003}). However, in our case, it is 
the vertical perturbation that couples with the rotation, playing an important
role to reveal (faster) instability. Our vertical perturbation is, therefore, 
equivalent to the non-axisymmetric perturbations 
required for PPI.

Now importantly the sign of the terms involved with
$q$ (but not $q$ itself) in the equations can be both positive and negative. 
The negative sign implies the opposite 
sense of rotation as compared to the positive sign. FIG.~\ref{fig:Poi_rot_pos_neg_q_3d_pure_Re_1000_1200} shows that the 
eigenspectra for plane Poiseuille flow in the presence of rotation with either
of the orientations for a vertical perturbation.
We notice that the eigenspectra are identical for both the orientations of 
Keplerian rotation for a 
fixed $Re_{\alpha}$. From equation (\ref{eq:plane_Poi_disp_rel}) and FIG.~\ref{fig:stability}, it is obvious that the
negative rotational effect (when the term involved with $1/q$ is negative) gets nullified by the negativity of 
$\mathcal{X}$, which is equivalent to the positive rotational effect in the positive 
$\mathcal{X}$ region. Hence, for the entire zone of $\mathcal{X}$, the net effect appears
to be unchanged in either of the orientations of rotation. This explains why eigenspectra are independent of the orientation of
rotation for a vertical perturbation. Nevertheless, for threedimensional perturbation also, this conclusion is true. 
The reason is the following. In order to obtain the eigenspectra, we should 
have the secular determinant corresponding to the operator 
$\mathcal{L}$ in equation ({\ref{eq:the_L_matrix})}. The information of rotation enters into the picture by 
$\mathcal{L}_{12}$ and $\mathcal{L}_{21}$. In that secular equation, $\mathcal{L}_{12}$ and $\mathcal{L}_{21}$ appear 
as multiplication between themselves. More interestingly, we notice that
\begin{equation}
 \mathcal{L}_{12} \mathcal{L}_{21} = \left(\frac{\mathcal{X}}{q}-\frac{1}{q^2}\right) 4k_z^2 (\mathcal{D}^2-k^2)^{-1}.
 \label{eq:L12_mul_L21}
\end{equation}
$\mathcal{L}_{12} \mathcal{L}_{21}$, therefore, does not depend on the orientation of
rotation because of the presence of $\mathcal{X}$ which spans from $-1$ to $+1$.

We also can obtain the domain of $q$ which could give rise to instability depending on other parameters. 
From equation (\ref{eq:plane_Poi_disp_rel}), the first condition for instability, irrespective of the orientation of rotation, is
\begin{equation}
 \frac{\mathcal{X}}{q}-\frac{1}{q^2} > 0,
 \label{eq:cond_for_instability}
\end{equation}
or in other words 
\begin{equation}
 \mathcal{X} > \frac{1}{q}.
 \label{eq:least_bound_on_q}
\end{equation}
For the flows with $q<1$ (when $q$ is positive for the present purpose), the above condition violates. Hence, our primary
domain of $q$ for the plausible unstable flows 
in the present context is $q < \infty$ excluding the domain $q \in [0,1]$. 
However, in the present context the domain of interest
is $q\in [1,2]$ and $q\rightarrow\infty$ (conventional plane Couette 
flow without rotation).

FIG.~\ref{fig:Poi_rot_two_q_3d_pure_Re_1000_1200} describes the eigenspectra for plane Poiseuille flow in the presence of 
rotation for a vertical perturbation to capture two phenomena. In one 
hand, it shows that for a fixed $q$, increment of $Re_{\alpha}$ increases the growth rates of a vertical perturbation.
On the other hand, it also depicts that the increment of $q$ for a fixed $Re_{\alpha}$, increases the growth rates of a 
vertical perturbation. However, the latter has more stronger effect than the earlier. 
This is because when we observe the 
flow for a fixed $q$ but at different $Re_{\alpha}$'s, we observe the same flow at different levels of initial velocity 
(lower $Re_{\alpha}$ corresponds to a more streamline flow). 
However, for different $q$'s, we altogether study the different flows, when the stronger rotation 
is more prone to instability.

FIG.~\ref{fig:poi_sigma_i_max_vs_q} describes the variation of maximum growth rate ($\sigma_{i, max}$) 
as a function of $q$ for vertical perturbation with $Re_{\alpha} = 1500$. Here the growth rates are maximized over the 
wavenumbers, $k_z$, i.e., we consider those $k_z$'s, which give rise to the maximum growth rate corresponding to 
each $q$. FIG.~\ref{fig:poi_sigma_i_max_vs_q} further shows that $\sigma_{i, max}$ increases with increasing $q$, which can 
be understood qualitatively from FIG.~\ref{fig:stability} and equation (\ref{eq:plane_Poi_disp_rel}).

FIG.~\ref{fig:Poi_rot_kep_3d_diff_ky_Re_1200} describes eigenspectra of plane Poiseuille flow in the presence of rotation 
for five different $k_y$. Interestingly, here we 
notice that as $k_y$ increases (i.e. the perturbation becomes more threedimensional from purely vertical in nature), the 
flow becomes more and more stabilized, or the unstable flow becomes stable. As we have already mentioned earlier that 
plane 
Poiseuille flow becomes unstable at a $Re=5772.22$ for planer (i.e. $k_z = 0.0$) perturbation. However, we have seen in 
the previous subsections and also we shall discuss
in \S\ref{sec:Discussion} that rotational effect makes the flow unstable at a 
$Re$ which is about two orders of magnitude 
lesser than that obtained based on a planer perturbation. For plane Poiseuille flow, therefore, the rotational effect (or 
the corresponding 
Coriolis effect) is more prone to lead to instability than that from 
Tollmien-Schlichting waves (\citealt{Alfredsson_1989}),
which are the corresponding planer perturbation modes at the critical $Re$.
See further \S\ref{sec:pert_ana_cpf} to understand other detailed physics behind
the eigenspectra.

\section{Perturbation analysis to rotating Couette-Poiseuille flow}
\label{sec:cou-poi}

\subsection{The formulation of dimensionless background flow}
\label{sec:cou_poi_bgf_dim_less}

As shown in \S\ref{sec:Background flow}, plane Couette flow in the presence of external force develops
a nonlinear shear in additional to its background linear shear, as shown by equation (\ref{eq:cou-PPF}). 
This is called Couette-Poiseuille flow. Of course in a suitable coordinate frame, plane Couette-Poiseuille flow will 
turn out to be plane Poiseuille flow as shown in equation (\ref{eq:PPF_reduction}). Nevertheless, as shown by FIG. 
\ref{fig:mod_base_flow}, depending upon the flow parameters, i.e. strength of force and background velocity, the domain 
of plane Poiseuille flow, more precisely plane Couette-Poiseuille flow, varies which further affects the flow behavior 
under perturbation. Here, we explore Couette-Poiseuille flow under various flow parameters. We, therefore, 
have to make the equation (\ref{eq:cou-PPF}) dimensionless. In dimensionless units, equation (\ref{eq:cou-PPF}) turns out 
to be 
\begin{equation}
 U = \xi (1-x^2) - x,
 \label{eq:cou_poi_nondim}
\end{equation}
where $\xi = KL^2/2U_0 = \Gamma_YL^2/2\nu U_0$ and $x = X/L$.
The background velocity vector, therefore, is $\textbf{U} = (0, U, 0)$.

To examine the stability of the background flow with velocity $\bfu$ within the domain,  
$x\in [-1,1]$, in a rotating frame at $R_0$, we consider 
the same prescription of angular velocity of the rotating frame 
as chosen in \S\ref{sec:Defining new Reynolds number}, given by 
$\bm{\omega} = (0,0,\Omega_0)$, $\Omega=\Omega_0(R/R_0)^{-q}$, and
$\Omega_0 = U_0/qL$. Although the similar kind of flow was explored by \cite{Balakumar}, they did not 
consider the effect of rotation which is crucial for astrophysical bodies particularly for accretion disks. Following the 
same procedure as done in \S\ref{sec:Defining new Reynolds number}, particularly from equations 
(\ref{eq:NS_rot}) to (\ref{eq:Re_alpha}), we can redefine the Reynolds number corresponding to the flow as 
\begin{equation}
 Re = \frac{U_0L}{\nu},
 \label{eq:Re}
\end{equation}
and $\bm{\Gamma}_{cp} = \bm{\Gamma}L/U_0^2$.
We can also rewrite $\xi$ as 
\begin{equation}
\xi = \Gamma_Y Re L/2U_0^2.
\label{eq:xi}
\end{equation}

\subsection{The perturbation analysis}
\label{sec:pert_ana_cpf}
To perform a perturbation analysis for the background flow 
described by equation (\ref{eq:cou_poi_nondim}), we follow 
the same procedure
described in \S\ref{sec:Perturbation analysis} and we obtain the corresponding Orr-Sommerfeld and Squire equations 
similar to equations (\ref{eq:Orr_sommerfeld}) and (\ref{eq:Squire}), given by
\begin{eqnarray}
 \begin{split}
  \left(\frac{\partial}{\partial t} + U\frac{\partial}{\partial y} \right)\nabla^2u - 
U''\frac{\partial u}{\partial y}+\frac{2}{q}\frac{\partial \zeta}{\partial z} \\- 
\frac{1}{Re}\nabla^4u  
= 0,\\
\label{eq:Orr_sommerfeld_CP}
 \end{split}
\end{eqnarray}
and 
\begin{eqnarray}
 \begin{split}
\left(\frac{\partial}{\partial t} + U\frac{\partial}{\partial y} \right)\zeta - U'\frac{\partial 
u}{\partial z}-\frac{2}{q}\frac{\partial u}{\partial z} \\- \frac{1}{Re}\nabla^2\zeta  = 0,
\label{eq:Squire_CP}
 \end{split}
\end{eqnarray}
where prime denotes differentiation w.r.to $x$.
The corresponding no-slip boundary conditions are: 
$u = v = w = 0$ at the two boundaries $x=\pm1$, or equivalently $u = \frac{\partial u}{\partial 
x} = \zeta = 0$ at $x=\pm1$ (see \citealt{man_2005, Ghosh_2021}).
We, then, substitute solution forms given by equations (\ref{eq:trail_u}) and (\ref{eq:trail_zeta}),  
but replacing $\mathcal{X}$ by $x$, in equations (\ref{eq:Orr_sommerfeld_CP}) and (\ref{eq:Squire_CP}), and 
eventually obtain equation (\ref{eq:lin_Q_eq}) through equations (\ref{eq:linear_orr_sommerfeld_eq_recast}) and 
(\ref{eq:linear_squire_eq}), where 
$\mathcal{L}$ and the elements of $\mathcal{L}$ are given by equations (\ref{eq:the_L_matrix}) and
(\ref{eq:L_diff_elements}) but replacing $U_{\alpha}$, $Re_{\alpha}$, and $\mathcal{D} = \partial/\partial \mathcal{X}$ by
$U$, $Re$, and $\mathcal{D} = \partial/\partial x$ respectively.

To have a qualitative idea about the eigenspectra, the analytical exploration based on pure vertical 
perturbations, shown in \S\ref{sec:Pure vertical perturbation} for plane Poiseuille flow, is of great use.
 Replacing $U'_{\alpha Y}$ by $U'$ in equation (\ref{eq:gen_disp_rel}) for Couette-Poiseuille flow, the 
growth rate turns out to be
\begin{equation}
 \sigma_{CP} = -\frac{\gamma_{CP}}{2} \pm \sqrt{-\frac{4}{q^2} +\frac{4\xi x}{q}+\frac{2}{q}},
 \label{eq:disp_cp2}
\end{equation}
where $\gamma_{CP} = 2k_z^2/Re$. For the marginal instability, the discriminant in equation 
(\ref{eq:disp_cp2}) becomes zero and hence 
\begin{eqnarray}
q = \frac{2}{2\xi x +1}.
\label{eq:q_mar_ins}
\end{eqnarray}
If we consider $\xi = 0$ in the equation (\ref{eq:q_mar_ins}), we see that 
the marginal instability occurs 
at $q = 2$. However, for the concerned flow, this condition is relaxed by the presence of $\xi$. The constraint on $q$ 
can be drawn from the domain size, i.e. $|x|\leq1$. The restrictions on $q$ for marginal instability, therefore, are
\begin{equation}
 q>\frac{2}{2\xi+1}\ {\rm for}\ x<1,
 \label{eq:x_less_1}
\end{equation}
and 
\begin{equation}
 q<\frac{2}{1-2\xi}\ {\rm for}\ x>-1.
 \label{eq:x_greater_-1}
\end{equation}
Similarly, we can obtain the constraint on $\xi$ too. To have the instability, the discriminant in 
equation (\ref{eq:disp_cp2}) has to follow the condition given by
\begin{equation}
-\frac{4}{q^2} +\frac{4\xi x}{q}+\frac{2}{q}\geq 0
\label{eq:cond_on_stability}
\end{equation}
and hence $\xi x\geq1/q-1/2$. The constraint on $\xi$, therefore, is given by
\begin{equation}
 \xi >\frac{1}{q} - \frac{1}{2}.
 \label{eq:cond_on_xi}
\end{equation}
For Keplerian flow, i.e. $q = 1.5$, $\xi>0.167$. 

\begin{figure}
\includegraphics[width=\columnwidth]{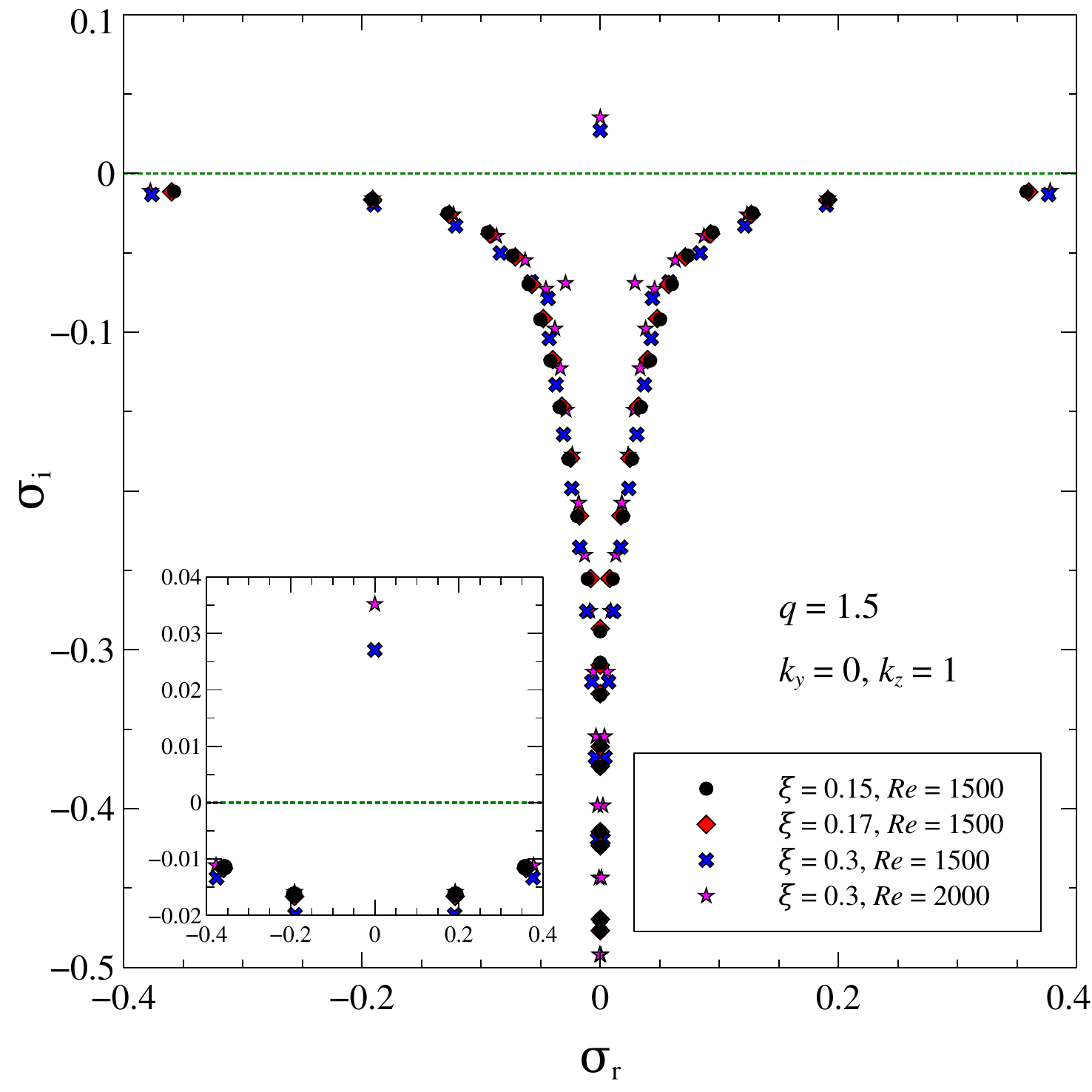}	
  \caption{Eigenspectra for Couette-Poiseuille flow, described by equation 
(\ref{eq:cou_poi_nondim}), in the presence of Keplerian rotation ($q = 1.5$) for vertical perturbation with $k_y = 0$ and 
$k_z = 1$ for different $Re$ and $\xi$.}
\label{fig:eval_cou_poi_re_1500_2000_ky_0_kz_1_q_kep_aa_diff}
\end{figure}

FIG.~\ref{fig:eval_cou_poi_re_1500_2000_ky_0_kz_1_q_kep_aa_diff} however describes the exact eigenspectra for 
Couette-Poiseuille flow in the presence of Keplerian rotation ($q = 1.5$) for vertical perturbation with $k_y = 0$ and 
$k_z = 1$ for several $\xi$ and $Re$. As the figure shows, for $\xi = 0.3$, the 
system is unstable, while for $\xi = 0.15$ and $0.17$, the system is stable. 
This is to remember that the apparent discrepancy between the results based on
equation (\ref{eq:cond_on_xi})
and FIG.~\ref{fig:eval_cou_poi_re_1500_2000_ky_0_kz_1_q_kep_aa_diff} is due
the inexact nature of eigenvalue given by equation (\ref{eq:disp_cp2}).
Equation (\ref{eq:disp_cp2}) corresponds to a qualitative 
description of the vertical perturbation for Couette-Poiseuille flow. While deriving equation (\ref{eq:disp_cp2}), we have 
considered the perturbation to be the function of $z$ only. 
However, in reality in order to obtain the eigenvalues (here for vertical 
perturbation), the perturbation has to be the function of both $x$ and $z$,
as indeed was considered in order to obtain eigenspectra given by 
FIG.~\ref{fig:eval_cou_poi_re_1500_2000_ky_0_kz_1_q_kep_aa_diff}. 
This leads to a differential equation of $x$, which we solve 
with no-slip boundary condition to have the eigenspectra.
Hence, while equation (\ref{eq:disp_cp2}) 
and the conditions derived from it give the qualitative idea for the 
eigenspectra, those do not provide 
the exact information.
In FIG.~\ref{fig:eval_cou_poi_re_1500_2000_ky_0_kz_1_q_kep_aa_diff}, we also 
notice that for $\xi = 0.3$, the system is more unstable for $Re = 2000$ than $Re = 1500$. It is quite easy to 
understand from equation (\ref{eq:disp_cp2}) that keeping other parameters, 
i.e., $q$, $\xi$, and $k_z$, fixed, if we 
increase $Re$, $\gamma_{CP}$ decreases, and hence $\sigma_{CP}$ increases.	

\begin{figure}
\includegraphics[width=\columnwidth]{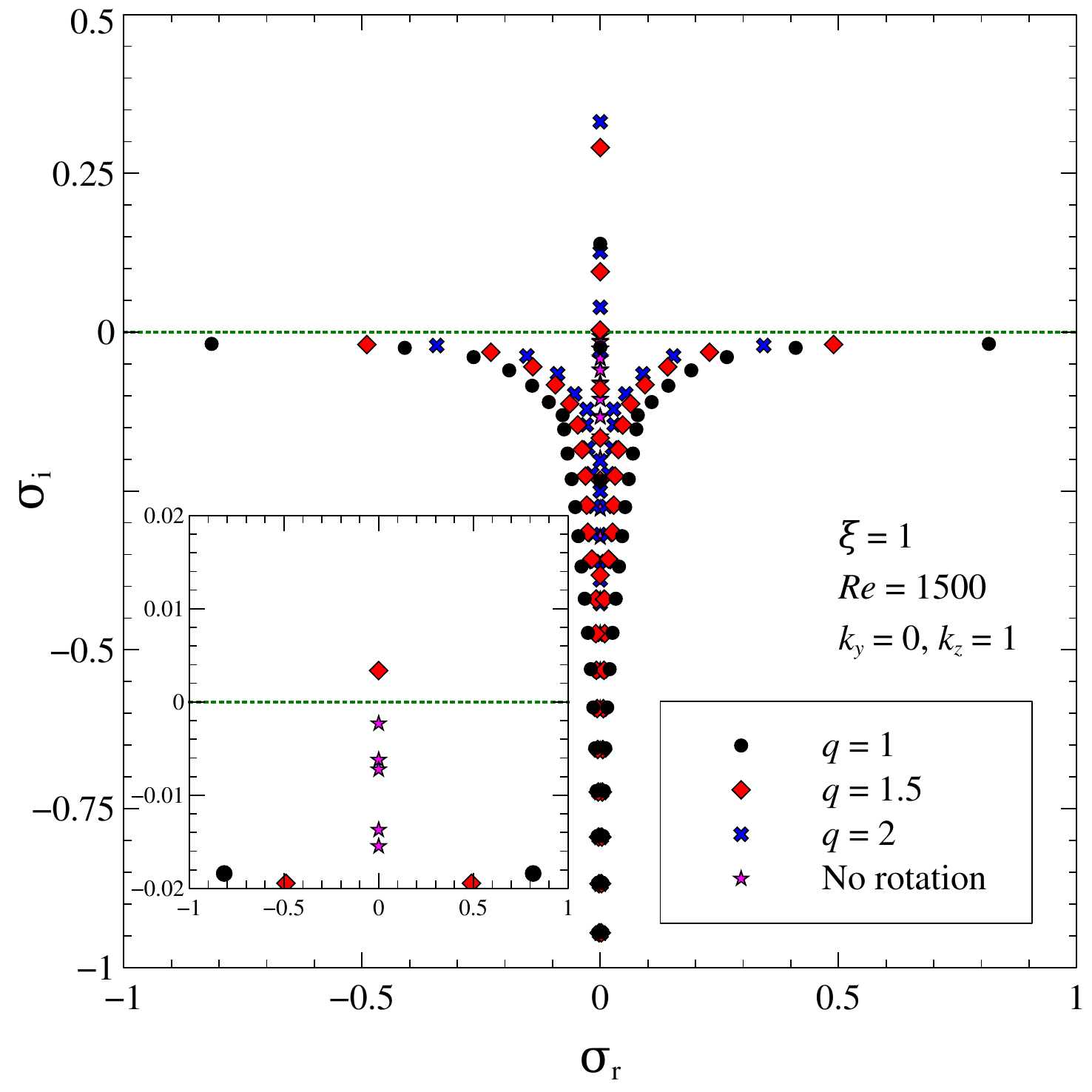}	
  \caption{Eigenspectra for Couette-Poiseuille flow, described by equation 
(\ref{eq:cou_poi_nondim}), in the presence of rotation for vertical perturbation with $k_y = 0$ and $k_z = 1$ for 
different rotation parameter $(q)$,  $Re = 1500$ and $\xi = 1$.}
\label{fig:eval_cou_poi_re_1500_ky_0_kz_1_q_diff_aa_1}
\end{figure}

\begin{figure}
\includegraphics[width=\columnwidth]{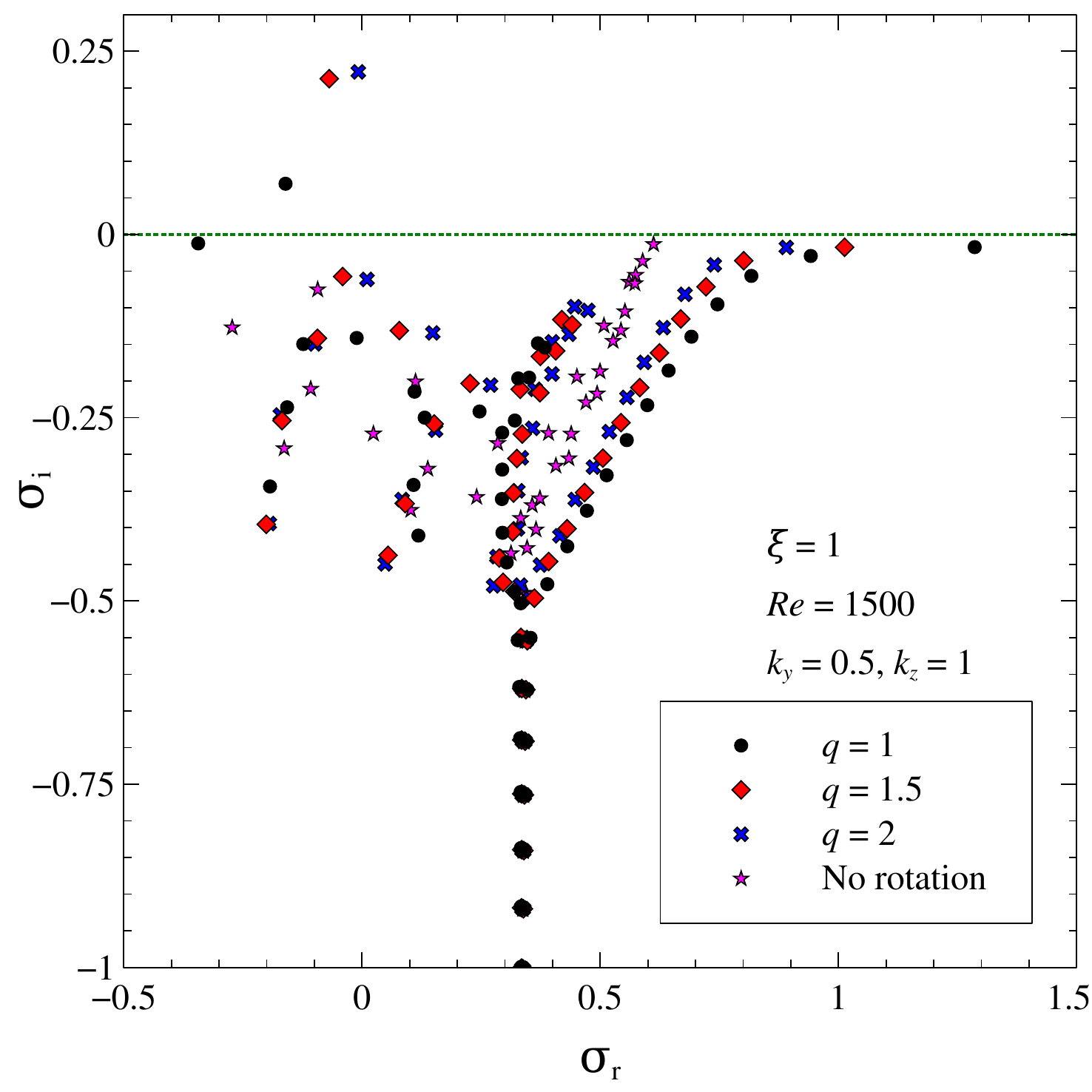}	
  \caption{Eigenspectra for Couette-Poiseuille flow, described by equation 
(\ref{eq:cou_poi_nondim}), in the presence of rotation for threedimensional perturbation with $k_y = 0.5$ and $k_z = 1$ 
for different rotation parameter $(q)$,  $Re = 1500$ and $\xi = 1$.}
\label{fig:eval_cou_poi_re_1500_ky_5_10_-1_kz_1_q_diff_aa_1}
\end{figure}

\begin{figure}
	\includegraphics[width=\columnwidth]{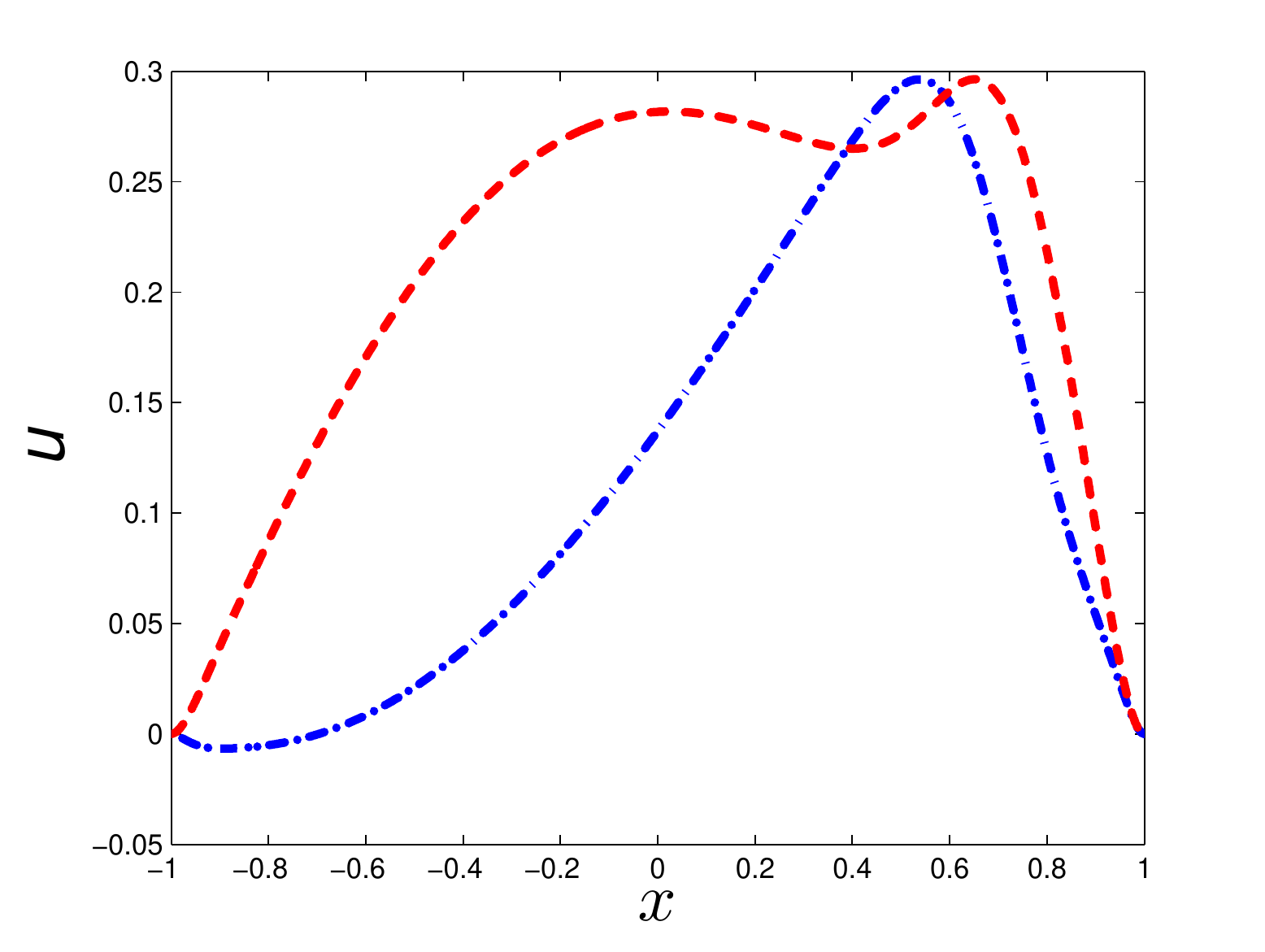} 
  \caption{Velocity eigenfunction for the most unstable mode corresponding to linearized 
Couette-Poiseuille flow 
in the presence of Keplerian rotation ($q=1.5$) of the box for $Re = 3000$ with $k_y = 0.5$ and $k_z = 1$. 
Dot-dashed and dashed lines indicate, respectively, the real and imaginary parts of $u$.}
\label{fig:evec_cou_poi_rot_3d_ky_point5_re_3000_kep_velo_xi_1}
\end{figure}

FIGs.~\ref{fig:eval_cou_poi_re_1500_ky_0_kz_1_q_diff_aa_1} 
and \ref{fig:eval_cou_poi_re_1500_ky_5_10_-1_kz_1_q_diff_aa_1} describe the eigenspectra for Couette-Poiseuille flow in 
the presence of rotation with different rotation parameters for vertical and threedimensional perturbations, 
respectively. For both cases, we see that as the rotation parameter
increases, the system becomes more unstable. If there is no rotation in the system, it is stable for the parameters 
considered in these two cases. FIG.~\ref{fig:evec_cou_poi_rot_3d_ky_point5_re_3000_kep_velo_xi_1} 
depicts an example for the nature of velocity eigenfunction, which is given 
for the most unstable mode corresponding to Couette-Poiseuille flow with $Re = 3000$, 
$k_y = 0.5$, $k_z = 1$, and $q=1.5$. According to \cite{Kersale2004}, these are body modes.

\begin{figure}
\includegraphics[width=\columnwidth]{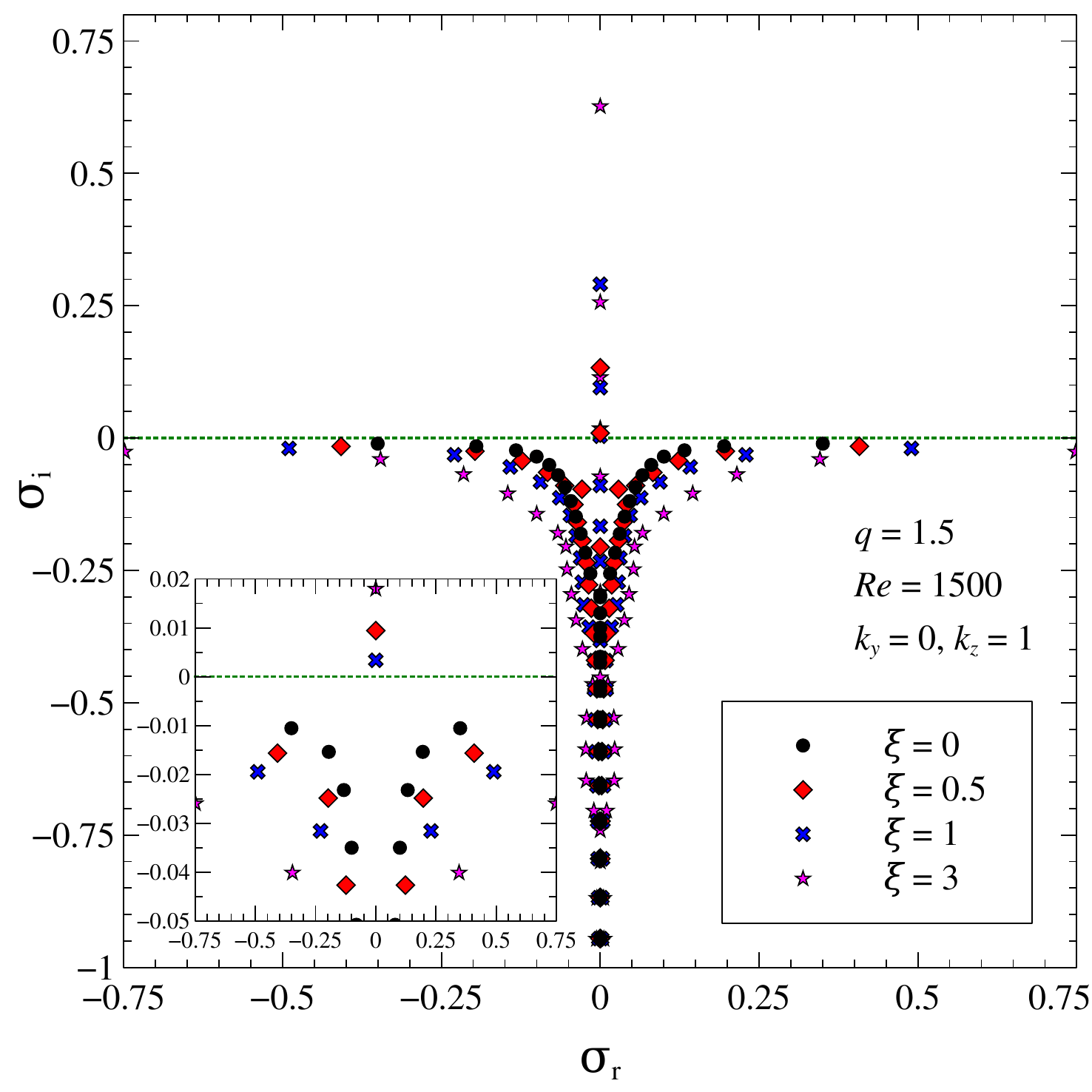}	
  \caption{Eigenspectra for Couette-Poiseuille flow, described by equation 
(\ref{eq:cou_poi_nondim}), in the presence of Keplerian rotation $(q = 1.5)$ for vertical perturbation with $k_y = 0$ and 
$k_z = 1$, $Re = 1500$ and different $\xi$.}
\label{fig:eval_cou_poi_re_1500_ky_0_kz_1_q_kep_aa_diff}
\end{figure}

\begin{figure}
\includegraphics[width=\columnwidth]{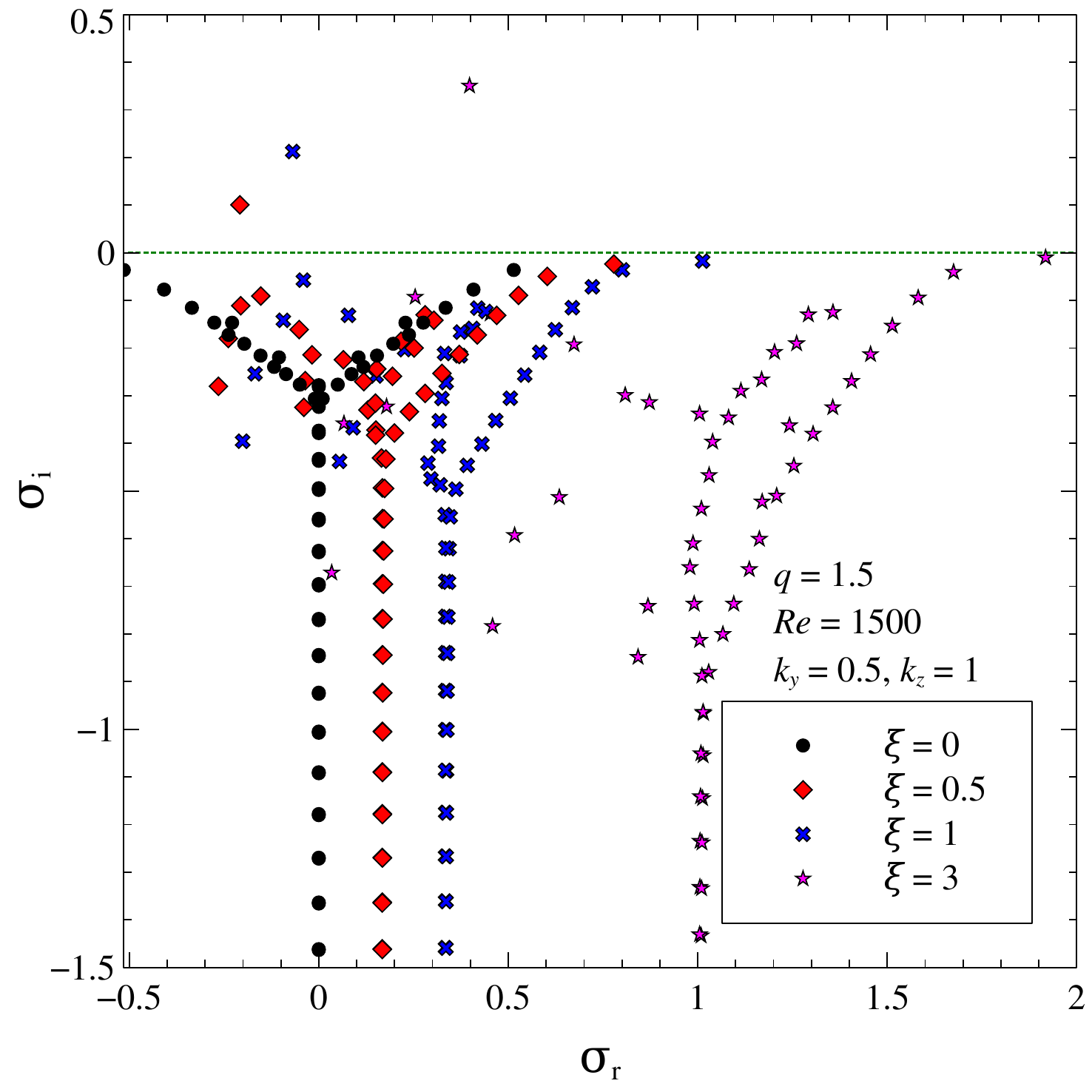}	
  \caption{Eigenspectra for Couette-Poiseuille flow, described by equation 
(\ref{eq:cou_poi_nondim}), in the presence of Keplerian rotation $(q = 1.5)$ for threedimensional perturbation with 
$k_y=0.5$ and $k_z = 1$, $Re = 1500$ and different $\xi$.}
\label{fig:eval_cou_poi_re_1500_ky_5_10_-1_kz_1_q_kep_aa_diff}
\end{figure}

FIGs.~\ref{fig:eval_cou_poi_re_1500_ky_0_kz_1_q_kep_aa_diff} 
and \ref{fig:eval_cou_poi_re_1500_ky_5_10_-1_kz_1_q_kep_aa_diff} describe the eigenspectra for Couette-Poiseuille flow in 
the presence of rotation for vertical and threedimensional perturbations, respectively, for different $\xi$. For both 
cases, we notice that as $\xi$ increases, the system becomes more 
unstable. From equation (\ref{eq:disp_cp2}), we can qualitatively explain this behavior. The discriminant in equation 
(\ref{eq:disp_cp2}) and hence $\sigma_{CP}$ increase as $\xi$ increases for a
fixed $q$.

\begin{figure}
\includegraphics[width=\columnwidth]{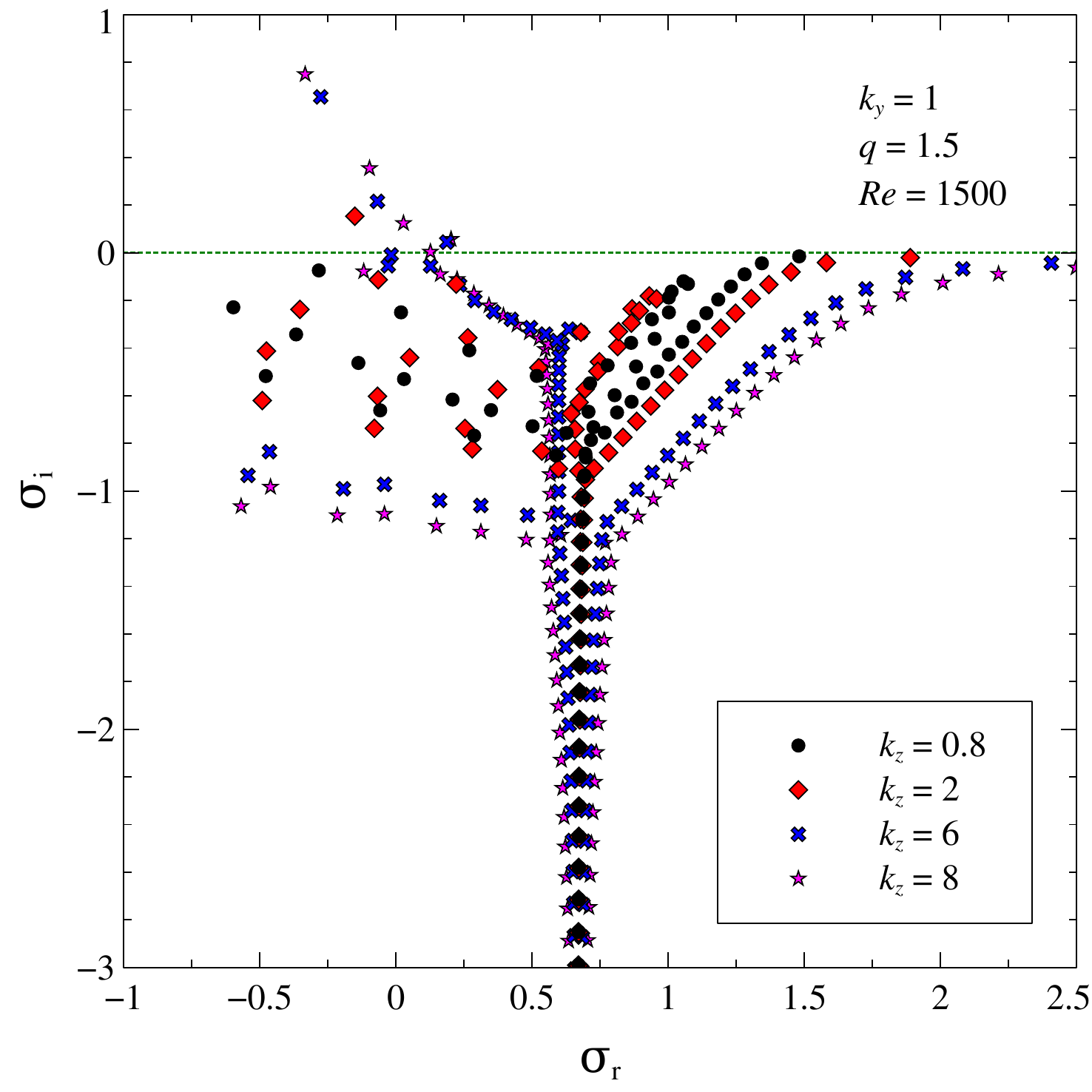}	
  \caption{Eigenspectra for Couette-Poiseuille flow, described by equation 
(\ref{eq:cou_poi_nondim}), in the presence of Keplerian rotation $(q = 1.5)$ for 
$k_y = 1$, different $k_z$, $Re = 1500$ and $\xi = 1$.}
\label{fig:eval_cou_poi_re_1500_ky_8_10_-1_8_kz_1_q_kep}
\end{figure}

\begin{figure}
\includegraphics[width=\columnwidth]{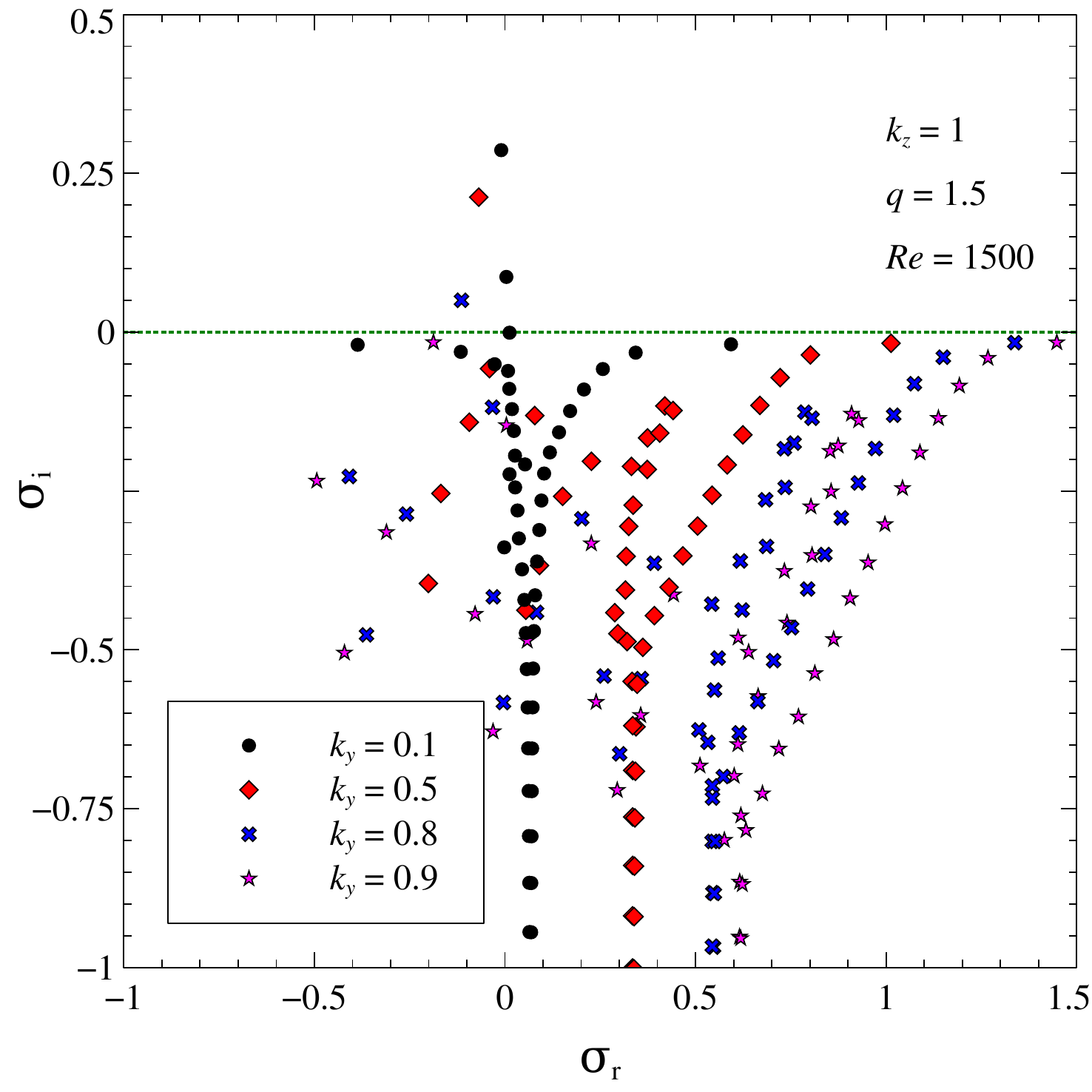}	
  \caption{Eigenspectra for Couette-Poiseuille flow, described by equation 
(\ref{eq:cou_poi_nondim}), in the presence of Keplerian rotation $(q = 1.5)$ for 
different $k_y$, $k_z = 1$, $Re = 1500$ and $\xi = 1$.}
\label{fig:eval_cou_poi_re_1500_ky_1_kz_1_9_10_-1_q_kep}
\end{figure}

FIG.~\ref{fig:eval_cou_poi_re_1500_ky_8_10_-1_8_kz_1_q_kep} describes the eigenspectra for 
Couette-Poiseuille flow in the presence of Keplerian rotation for for $k_{y} = 1$ and various $k_{z}$. We notice that as $k_{z}$ increases keeping other 
parameters fixed, the 
system becomes more unstable. From equation (\ref{eq:L_diff_elements}), it is 
clear that rotation (i.e. $q$) is coupled 
with $k_z$ and the shear velocity is coupled with $k_y$. As $k_{z}$ increases, 
the effect of rotation in the system, 
therefore, dominates over the effect of shear. It is now quite evident from 
the whole discussion that rotation and shear 
have opposite effects on the stability of the flow. Rotation tries to make the 
flow more unstable, while shear tries to 
stabilize it. It is also well-described in FIG.~\ref{fig:eval_cou_poi_re_1500_ky_1_kz_1_9_10_-1_q_kep}. It describes the 
eigenspectra for Couette-Poiseuille flow in the presence of Keplerian rotation for $k_{z} = 1$ and various $k_{y}$. 
As $k_y$ increases, the flow becomes more dominated by shear 
than rotation and hence making the flow more stable.

	Interestingly, it is seen from FIGs. \ref{fig:Poi_rot_kep_3d_diff_ky_Re_1200}, 
\ref{fig:eval_cou_poi_re_1500_ky_5_10_-1_kz_1_q_kep_aa_diff}, \ref{fig:eval_cou_poi_re_1500_ky_1_kz_1_9_10_-1_q_kep} 
that the entire eigenspectrum keeps shifting to the 
positive $\sigma_r$ direction with increasing $k_y$ or $\xi$. This can be 
qualitatively understood in the following analysis. Assuming approximate 
solutions for equations (\ref{eq:Orr_sommerfeld_CP}) and (\ref{eq:Squire_CP})
be $u,\zeta\sim \exp(\sigma_{CP3}t+\tilde{\textbf{k}}\cdot\textbf{r})$, where 
$\tilde{\textbf{k}}\equiv(k_x,k_y,k_z)$ and $\textbf{r}\equiv(x,y,z)$, we obtain
\begin{eqnarray}
	\sigma_r=\left(U+\frac{U^{''}}{2\tilde{k}^2}\right)k_y \pm 
f(k_x,k_y,k_z,U',U'', q),
	\label{sigcp3}
\end{eqnarray}
where the function $f(k_x,k_y,k_z,U',U'', q)$ is determined by the natures of 
background flow, perturbation and rotation, $\sigma_{CP3}=\sigma_i-i\sigma_r$, according to our convention of 
original exact solution given by equation (\ref{eq:gen_sol_Q}) and eigenspectra.
Note however that the solution in the $x$-direction in principle should not be
of the plane wave form as $U$ is a function of $x$, what indeed we do not choose
in order to compute eigenspectra. 
It is easy to check that the magnitude of $f(k_x,k_y,k_z,U',U'', q)$ is smaller than the first term in the parenthesis in 
equation 
(\ref{sigcp3}), and $\sigma_r$ increases with increasing
$k_y$ for the background flows and generally the parameters considered here.
Therefore, equation (\ref{sigcp3}) confirms that as $k_y$ increases, 
$\sigma_r$ increases, with a shift in the positive $\sigma_r$ direction, 
as seen in FIGs. \ref{fig:Poi_rot_kep_3d_diff_ky_Re_1200} and 
\ref{fig:eval_cou_poi_re_1500_ky_1_kz_1_9_10_-1_q_kep}. 
Similarly, with increasing $\xi$, Couette-Poiseuille flow tends to become pure Poiseuille flow and
$\sigma_r\rightarrow (1-x^2-1/\tilde{k}^2)\xi k_y$. Therefore, $\xi$ and $k_y$ play
interchangeable roles and, hence, with increasing $\xi$, $\sigma_r$ increases,
as seen in FIG. \ref{fig:eval_cou_poi_re_1500_ky_5_10_-1_kz_1_q_kep_aa_diff}.

\begin{figure}
\includegraphics[width=\columnwidth]{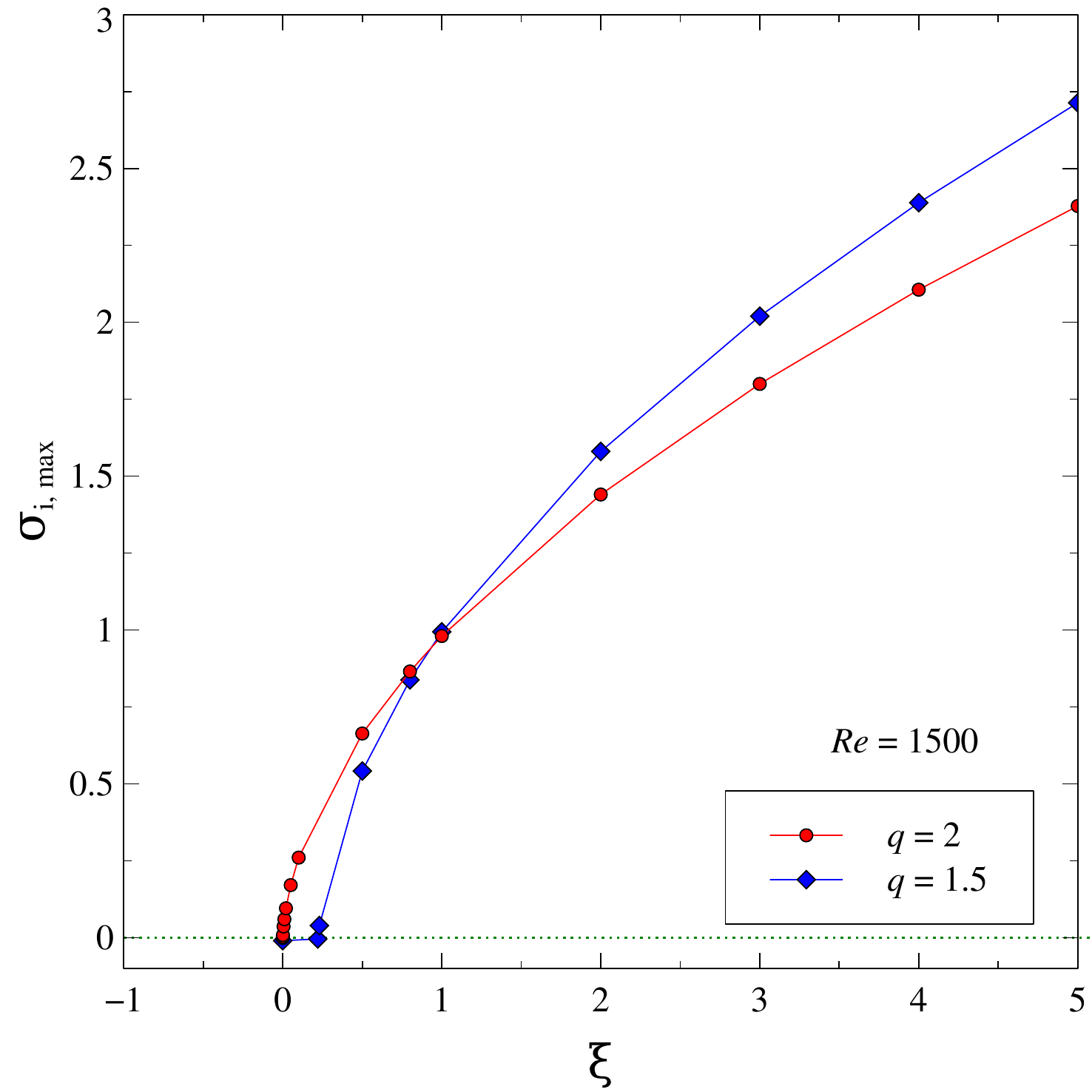}	
  \caption{Maximum growth rate ($\sigma_{i, max}$) as a function of $\xi$ for Couette-Poiseuille flow with
  $q = 1.5,2.0$ having vertical perturbation for $k_z$ maximizing $\sigma_{i, max}$ and $Re = 1500$.}
\label{fig:sigma_i_max_vs_xi}
\end{figure}

\begin{figure}
\includegraphics[width=\columnwidth]{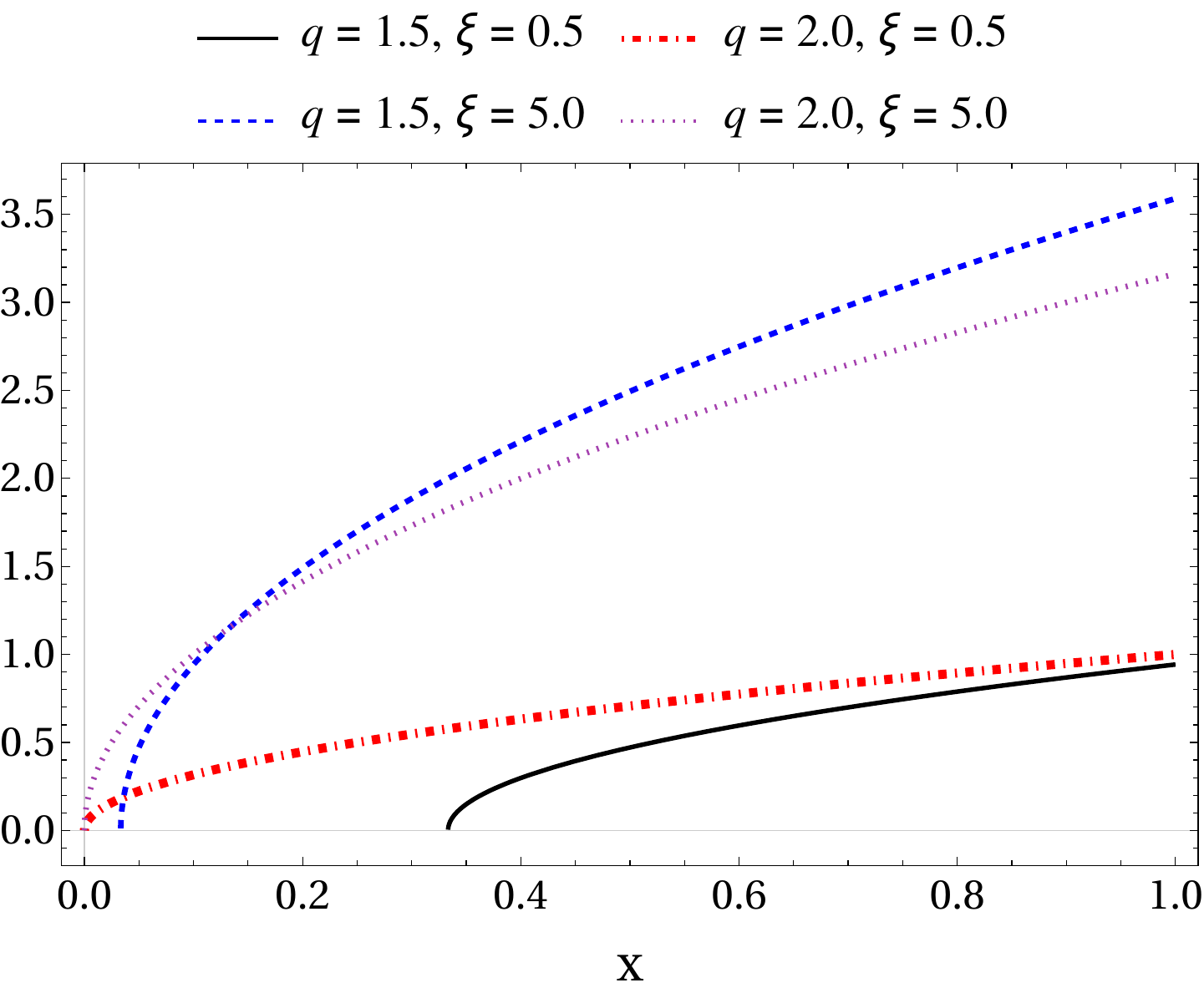}	
  \caption{Variation of $\sqrt{-4/q^2 +4\xi x/q+2/q}$ from equation (\ref{eq:disp_cp2}) as 
  a function of $x$ for several combinations of $q$ and $\xi$.}
\label{fig:stability_CPF_several_q_xi_combo}
\end{figure}

FIG.~\ref{fig:sigma_i_max_vs_xi} describes the maximum growth rate ($\sigma_{i, max}$) as a function of
$\xi$ for Couette-Poiseuille flow having a vertical perturbation with 
$k_z$ maximizing $\sigma_i$. It shows that $\sigma_{i, max}$ increases monotonically with $\xi$ for both the $q$'s.
It also shows that if $\xi\lesssim1$, $\sigma_{i, max}$ for $q = 2$ is larger than that for 
$q = 1.5$. However, the situation reverses for $\xi>1$. This phenomenon can be understood qualitatively from 
FIG.~\ref{fig:stability_CPF_several_q_xi_combo}, where we show the variation of $\sqrt{-4/q^2 +4\xi x/q+2/q}$ 
from equation (\ref{eq:disp_cp2}) as a function of $x$ for several combinations of $q$ and $\xi$.
We see that for $\xi = 0.5$, $\sqrt{-4/q^2 +4\xi x/q+2/q}$ is larger for $q = 2$ than that for $q = 1.5$. 
As a result, $\sigma_{CP}$ in equation (\ref{eq:disp_cp2}) becomes larger for $q = 2$ than that for $q = 1.5$. This explains
the behavior of $\sigma_{i, max}$ for $\xi\lesssim 1$ in the FIG.~\ref{fig:sigma_i_max_vs_xi}. Similarly, the explanation 
of larger $\sigma_{i, max}$ for $\xi>1$ in the FIG.~\ref{fig:sigma_i_max_vs_xi} can be extracted from the curves 
with $q = 1.5$ and $q = 2$ at $\xi = 5.0$, in the FIG.~\ref{fig:stability_CPF_several_q_xi_combo}. It is 
further verified from FIG~\ref{fig:sigma_i_max_vs_xi}
that below certain $\xi$, depending on $q$, flow becomes stable with negative 
$\sigma_{i, max}$.

\begin{figure}
\includegraphics[width=\columnwidth]{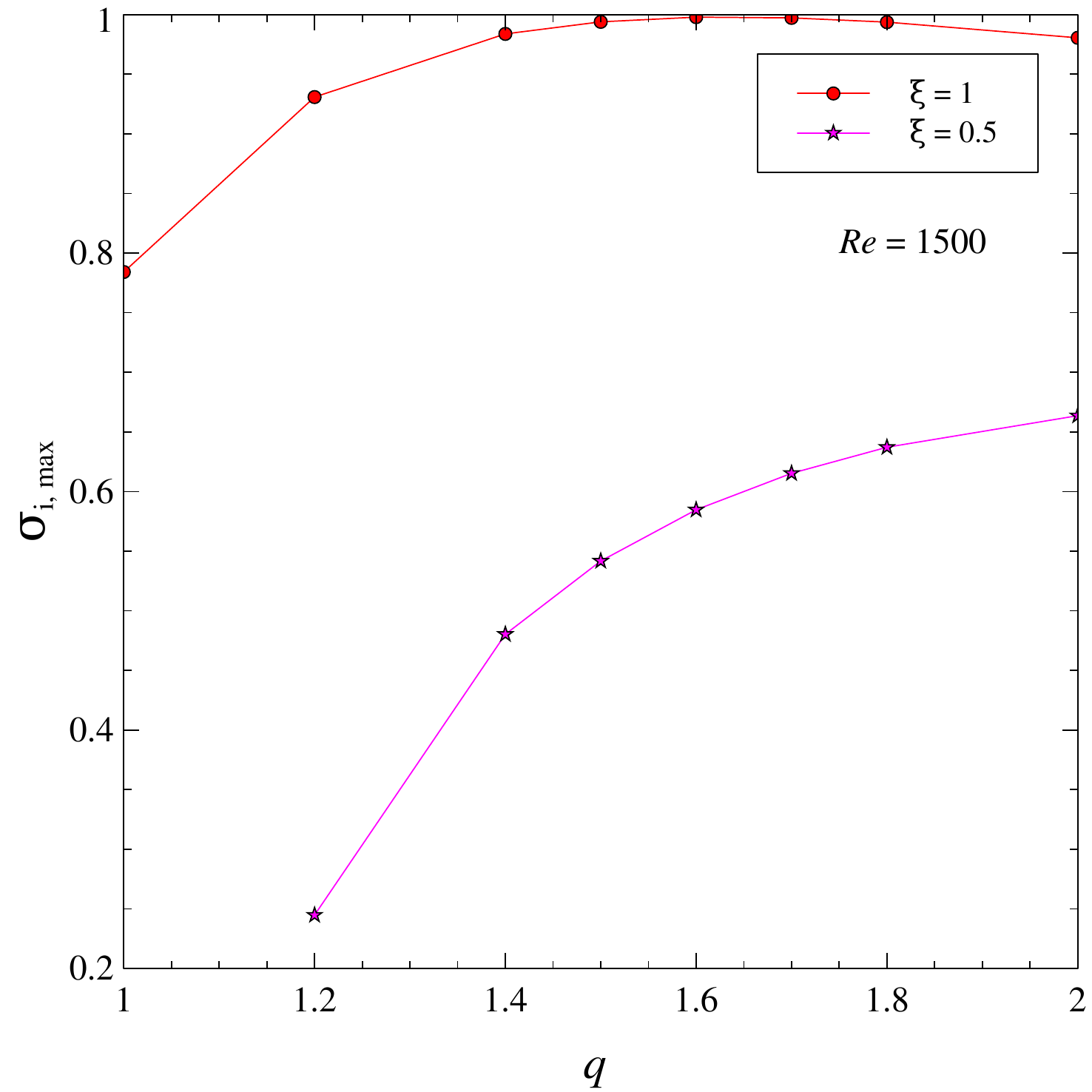}	
  \caption{Maximum growth rate ($\sigma_{i, max}$) as a function of $q$ for Couette-Poiseuille flow
  with $\xi = 0.5, 1.0$ having vertical perturbation for $k_z$ maximizing 
	$\sigma_{i, max}$ and $Re = 1500$.}
\label{fig:sigma_i_max_vs_q_2_xi}
\end{figure}

\begin{figure}
\includegraphics[width=\columnwidth]{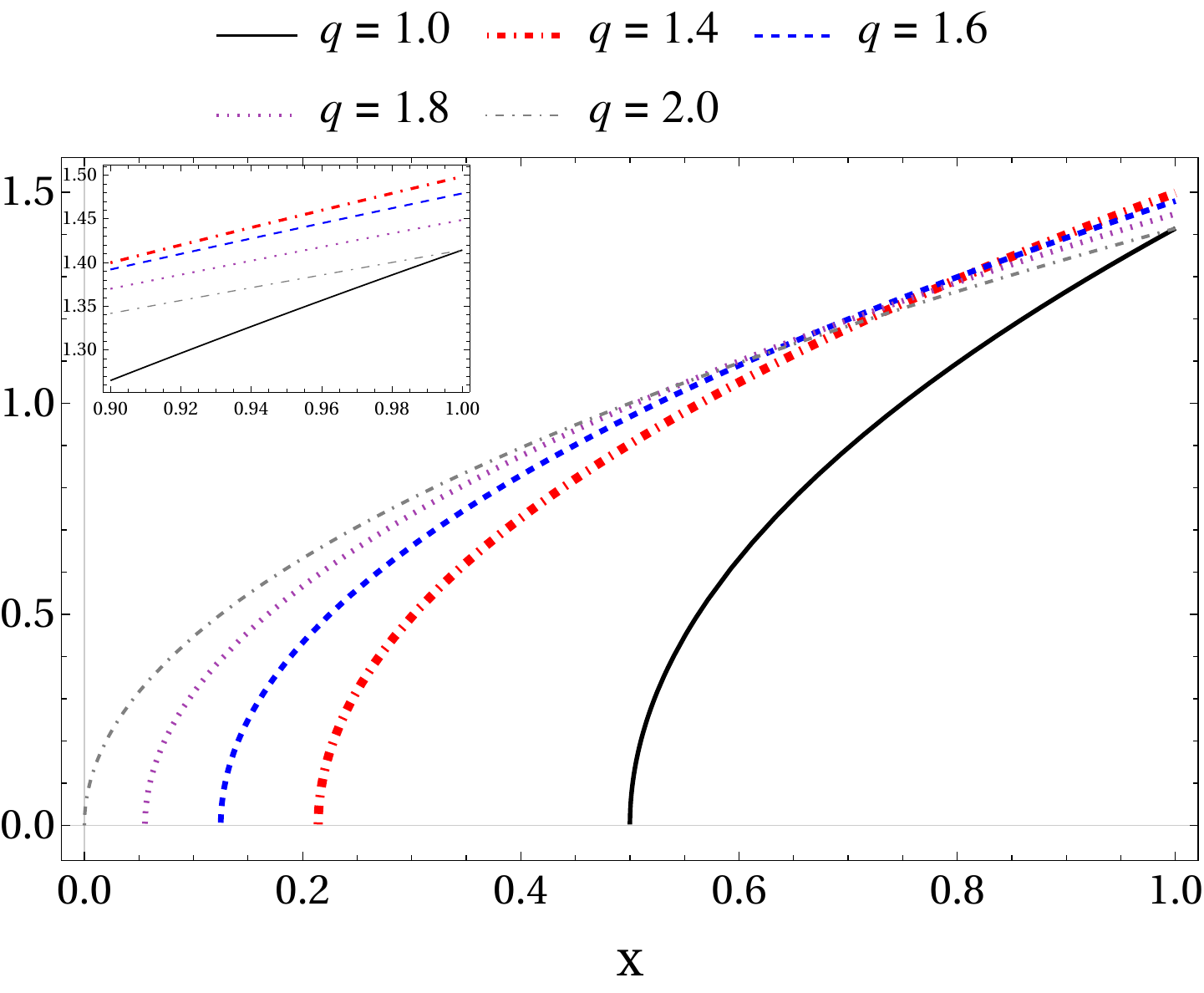}	
  \caption{Variation of $\sqrt{-4/q^2 +4\xi x/q+2/q}$ from equation (\ref{eq:disp_cp2}) as 
  a function of $x$ for $\xi = 1$ and for several $q$'s.}
\label{fig:stability_CPF_several_q}
\end{figure}

FIG.~\ref{fig:sigma_i_max_vs_q_2_xi} describes maximum growth rate as a function of
$q$ for Couette-Poiseuille flow with $\xi = 0.5,1.0$ and $Re = 1500$ for vertical perturbation with 
$k_z$ maximizing $\sigma_i$. It shows that $\sigma_{i, max}$ increases with
increasing $q$ for $\xi = 0.5$. However, 
for $\xi = 1$, $\sigma_{i, max}$ increases with increasing $q$ only up to 
$q \sim 1.6$, subsequently it decreases. This behavior can 
be qualitatively understood from FIG.~\ref{fig:stability_CPF_several_q}, where we show the variation of the discriminant in 
the equation (\ref{eq:disp_cp2}) as a function of $x$ with $\xi = 1$ and for $q = 1.0,\ 1.4,\ 1.6,\ 1.8\ \rm{and}\ 2.0$. 
It shows that the case of $q = 1.4$ gives rise to maximum discriminant. However, the 
maximum discriminants for $q=1$ and $2$ are almost the same and they are 
the least among all the $q$'s. Hence $\sigma_{CP}$ from 
equation (\ref{eq:disp_cp2}) will be the highest for $q = 1.4$ and lowest for $q = 1$ and $2$. This qualitative analysis, 
therefore, 
indicates that $\sigma_{i, max}$ is not expected to increase with the increase 
of $q$ throughout at any $\xi$.

\subsection{Viable magnitude of force}
\label{sec:Smallness of force}
From the bound on $\xi$ for Keplerian flow described above, we can have an estimation of the extra force, $\Gamma_Y$. 
From equation (\ref{eq:xi}), we have 
\begin{equation}
 U_0=\sqrt{\frac{\Gamma_Y L Re}{2\xi}}.
 \label{eq:U_0}
\end{equation}
The size of the shearing box, $L$, is $0.05R_s$ \citep{Nath_2015}, where $R_s = 2GM/c^2$ is the Schwarzschild radius 
for the central black hole of mass $M$ with $G$ and $c$ the gravitational constant and speed of light in 
free space, respectively. From \cite{Mukho_2013}, we get that for accretion disk, $Re\ga10^{14}$. Considering all these and 
the lowest bound on $\xi$, i.e. $\xi = 0.167$ for Keplerian disk,  we obtain
\begin{equation}
 U_0 = \sqrt{5\overline{m}\ \overline{n}\ \Gamma_Y}\times 10^9\ \rm{cm/sec} 
 \simeq \sqrt{\overline{m}\ \overline{n}\ \Gamma_Y}\frac{c}{10},
 \label{eq:U_0_with_values}
\end{equation}
where $Re = \overline{n} 10^{14}$, $M = \overline{m} M_\sun$ and $M_\sun$ is the mass of the Sun. Now at radius $R$, the 
speed would be 
\begin{equation}
 U_0=\sqrt{\frac{GM}{R}} = \frac{c}{\sqrt{2\rho}},
 \label{eq:U_0_grav}
\end{equation}
where, $R = \rho R_s$. If the fluid is at $100R_s$, then $U_0 = c/10\sqrt{2}$. From equations (\ref{eq:U_0_with_values}) and
(\ref{eq:U_0_grav}), we obtain
$\Gamma_Y = 0.5/\overline{m}\ \overline{n}\  \rm{cm/sec^2}$. This confirms that the extra force indeed is very small for 
the accretion disk around an astrophysical black hole whose $Re$ is huge. For example, a supermassive black hole of mass
$10^7M_\sun$ having accretion disk with $Re = 10^{22}$ leads to $\Gamma_Y = 5\times 10^{-16} \rm{cm/sec^2}$ which is too small
compared to the acceleration due to the gravity of the black hole at that position. This confirms that indeed a 
tiny $\Gamma_Y$, i.e., a very small effect of external force would make the flow unstable.

\section{Accuracy of Numerics}
\label{sec:Numerical technique}
\begin{figure}
\includegraphics[width=\columnwidth]{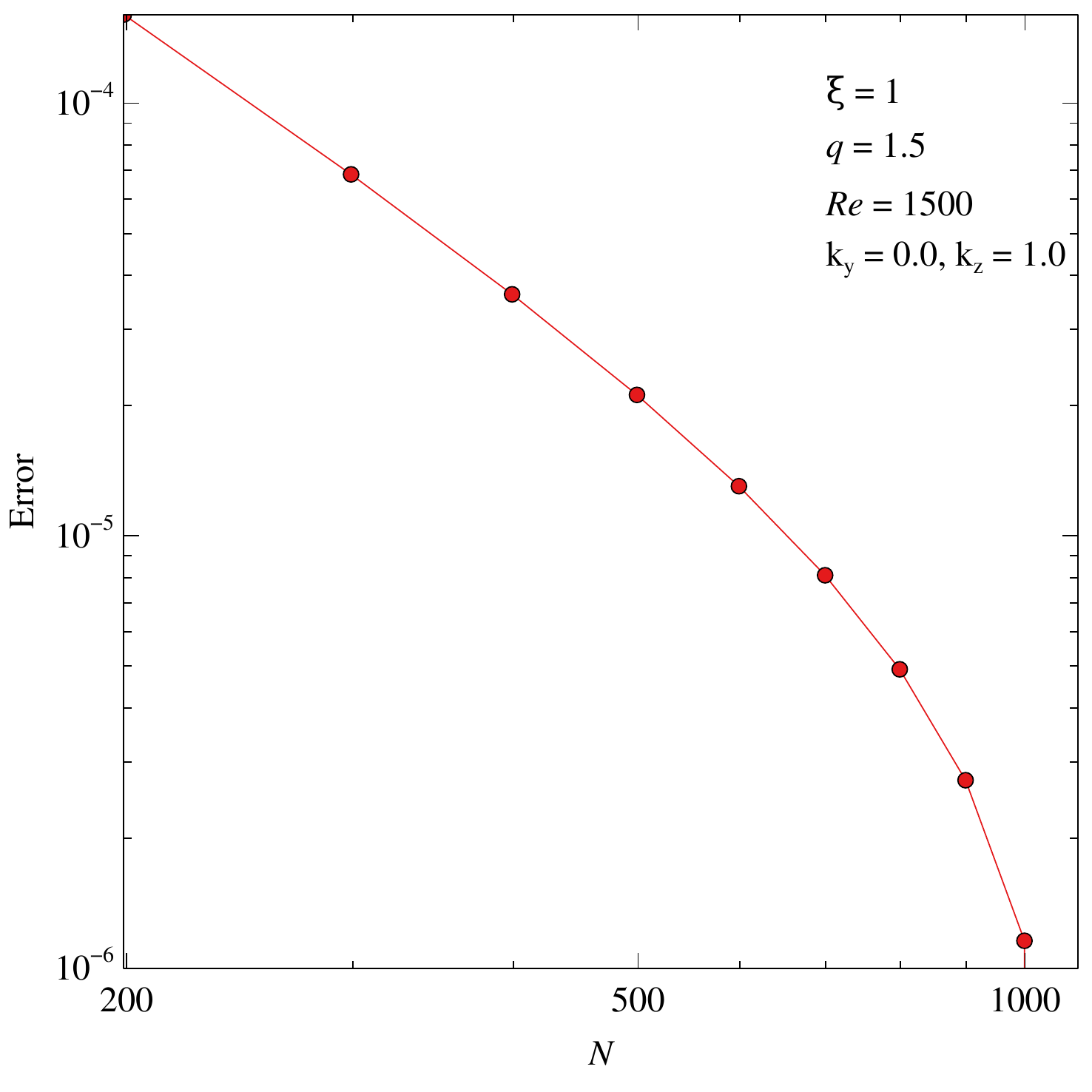}	   
  \caption{Error as a function of $N$ for Couette-Poiseuille flow with $\xi = 
1$, $q = 1.5$, $Re = 1500$, $k_y = 0$ and $k_z = 1.0$.}
\label{fig:max_eval_vs_N_cou_poi_Re_1500_kep_ver}
\end{figure}

Throughout the work, we have used the finite difference method to obtain the eigenspectra. We particularly have used 
the second order central difference method. Equation (\ref{eq:eival_eq}) is an eigenvalue equation, which is function $x$. To solve it 
numerically, we discretize the domain which ranges from $x = x_0 = -1$ to $x = x_f = 1$. In our calculation, we have 
divided the domain in $(N+1)$ segments, where the width of the each segment is defined as 
\begin{equation}
h = \frac{x_f - x_0}{N+1}.
\label{eq:def_h}
\end{equation}
For all the eigenspectra presented in this work, $N = 499$. Therefore, the dimension of $\mathcal{L}$ in 
equation (\ref{eq:eival_eq}) after using finite difference method is $2N\times2N$. 
To check the accuracy and the convergence of the eigenvalues for the chosen
matrix dimension, we show the variation 
of Error $= \sigma_{i, max}(N) - \sigma_{i, max}(N = 1099)$ as a function of $N$ in FIG.~\ref{fig:max_eval_vs_N_cou_poi_Re_1500_kep_ver} for 
a typical set of parameters. It confirms that chosen $N=499$ leads to the optimum numerical values of $\sigma_i$ which 
hardly changes with further increasing $N$. Infact, variation of $\sigma_{i, max}$ for $199\leq N \leq 1099$
is not more than $\sim10^{-4}$.

\begin{figure}
\includegraphics[width=\columnwidth]{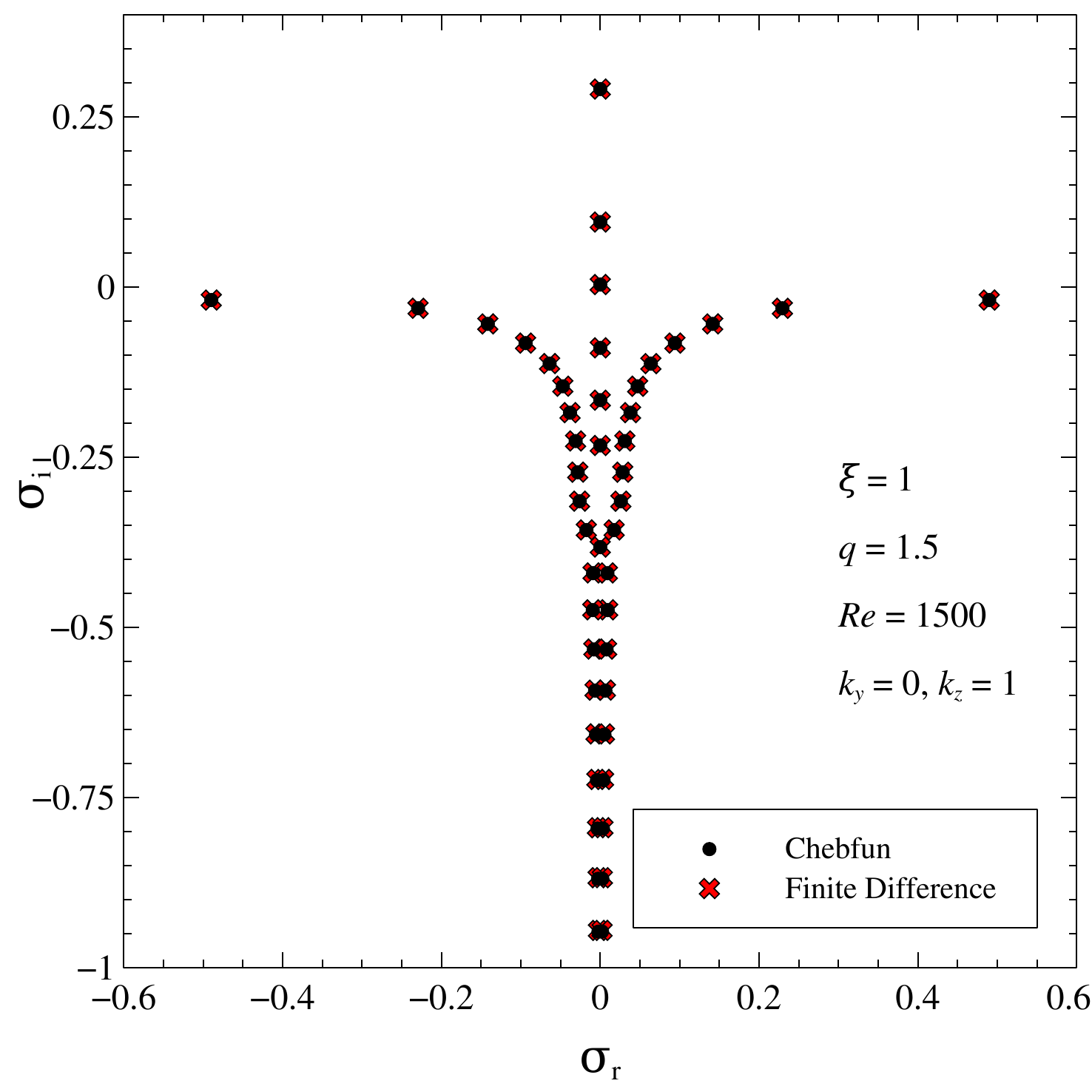}	
  \caption{Eigenspectra for Couette-Poiseuille flow with $\xi = 1$ in the presence of Keplerian rotation for 
$Re = 1500$, $k_y = 0$ and $k_z = 1$, obtained using Chebfun and finite difference methods.}
\label{fig:eval_cou_poi_re_1500_ky_0_kz_1_q_kep_aa_1_matlab_finite_diff}
\end{figure}

However, to check the accuracy of the eigenspectra, particularly the most unstable modes as these are the 
most important feature of this work, we have also verified the result of finite difference method with those obtained 
using Chebfun \citep{Chebfun}. FIG.~\ref{fig:eval_cou_poi_re_1500_ky_0_kz_1_q_kep_aa_1_matlab_finite_diff} 
demonstrates the eigenspectra for Couette-Poiseuille flow for a given set of
parameters. It confirms that both the eigenspectra match quite well with 
each other, hence the accuracy of our results.

\section{Discussion}
\label{sec:Discussion}
\begin{figure}
\includegraphics[width=\columnwidth]{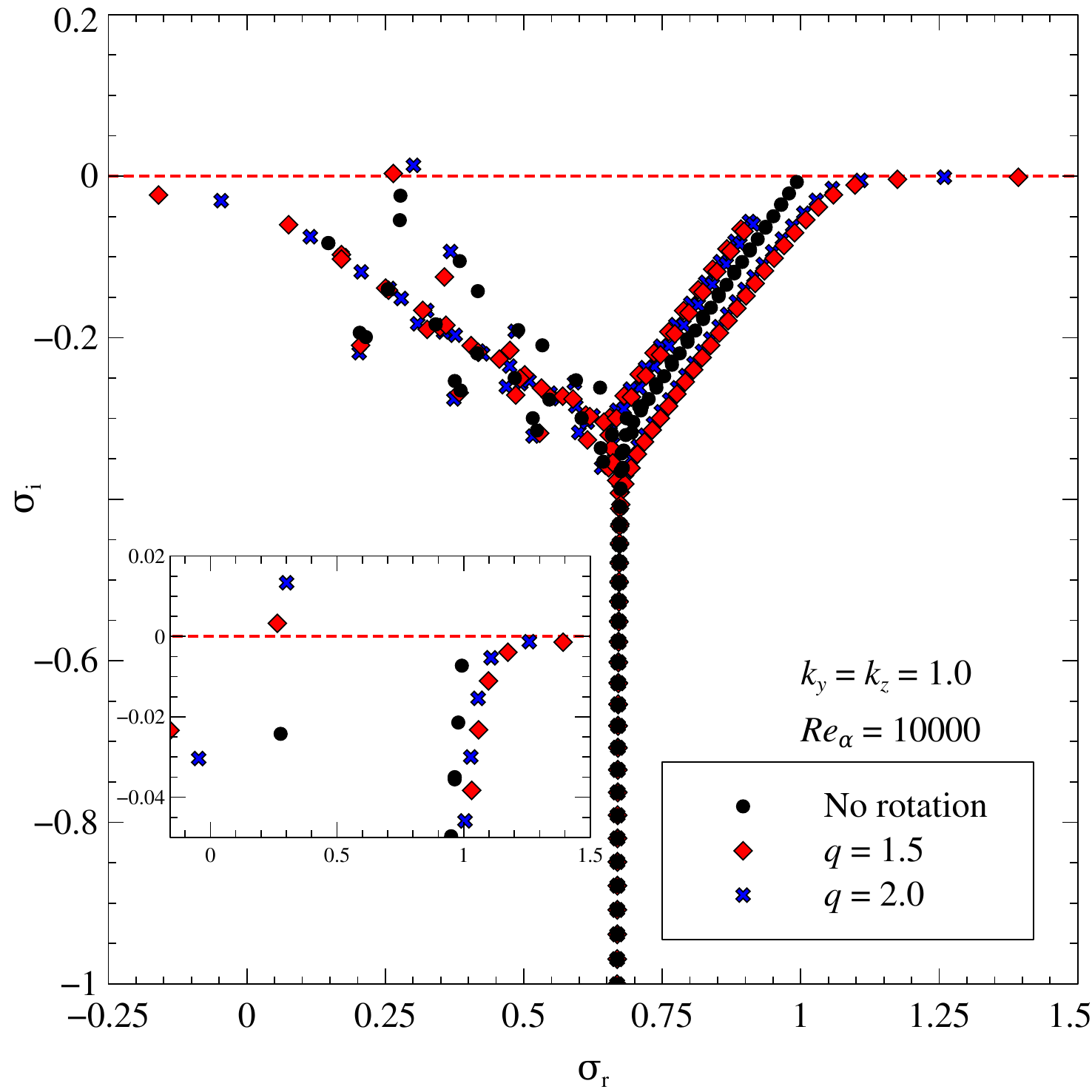}	
  \caption{Eigenspectra of linearized Poiseuille flow in the presence of rotation for threedimensional perturbation 
  with $k_y =k_z = 1$ for three different $q$ and $Re_{\alpha} = 10000$.}
\label{fig:Poi_rot_diff_q_3d_Re_10000}
\end{figure}

\begin{table}
	\caption{Critical values of Reynolds number ($Re_{\alpha, cric}$) and critical values of $k_z$ ($k_{z, cric}$) for
	three different rotation parameter $q$ for plane Poiseuille flow in the presence of rotation for vertical perturbation, i.e.
	$k_y = 0$.}
	\label{tab:table_cric}
	\begin{tabular}{lcccr} 
		\hline
		\hline
		$q$ &$Re_{\alpha, cric}$ &$k_{z, cric}$& \\
		\hline
		  2.0 & 137.2969 & 4.066 & \\ 
		 \hline
		  1.8 & 174.84 & 4.5762 & \\
		  \hline
		  1.5 & 327.58 & 6.1 & \\	
		  \hline
		\hline
	\end{tabular}
\end{table}

In the previous sections, we have observed that stability of rotating Poiseuille flow and Couette-Poiseuille flow greatly depends on $q$,
and also on the nature of perturbation. To make this statement
more concrete, we show in FIG.~\ref{fig:Poi_rot_diff_q_3d_Re_10000} the eigenspectra of plane Poiseuille 
flow in
the presence and absence of rotation for threedimensional perturbation with $k_y =k_z = 1$. Here, 
we notice that Poiseuille flow is stable even for $Re_{\alpha} = 10000$ with $k_y =k_z = 1$, when rotational 
effect has not been taken into account. On the contrary, when rotation is there, the flow becomes unstable, 
and as $q$ increases, 
the maximum growth rate increases for the same set of other parameters. We, therefore, argue that rotation makes 
plane Poiseuille flow unstable. On the other hand, rotation has an opposite effect on plane Couette flow. 
It is clear from
equation (\ref{eq:gen_disp_rel}) by substituting $U_{\alpha Y} = -\mathcal{X}$ and hence 
$U'_{\alpha Y} = -1$. Equation (\ref{eq:gen_disp_rel}) then becomes 
\begin{equation}
 \sigma = -\frac{\gamma}{2} \pm \frac{\sqrt{2}}{q}\sqrt{q-2},
 \label{eq:plane_cou_dis_rel}
\end{equation}
and hence
\begin{equation}
 \sigma = -\frac{\gamma}{2} \pm \frac{i}{\Omega q}\kappa,
 \label{eq:plane_cou_dis_rel_epi_freq}
\end{equation}
where $\kappa$ is the epicyclic frequency, given by $\kappa = \Omega\sqrt{2(2-q)}$.
Now for $q<2$, $\kappa$ is always a positive real number and, hence, plane Couette flow with rotational effect
is always stable as long as $q<2$. On the other hand, equation (\ref{eq:plane_Poi_disp_rel}) shows that even 
if $q<2$, plane Poiseuille flow with rotation becomes unstable in a particular domain of flow depending 
on $q$.

In \S\ref{sec:Dependence of rotation}, we have argued that as $q$ increases for a fixed $Re_{\alpha}$, the maximum growth 
rates increases. This statement quite matches with the literature, i.e. \citealt{Lezius_1976} (its figure 2), 
\citealt{Alfredsson_1989}. However, those authors considered `$Ro$' as the rotation parameter and it is inverse of $q$. 
Apart from that, the background flow considered by them is $6(\mathcal{X} - \mathcal{X}^2)$. This is the reason behind 
obtaining different critical $Re$ and wavevector in the present work than 
those by \cite{Lezius_1976} and \cite{Finlay_1990} (see their table 1). 
While they obtained critical $Re\sim$ 89 and 
critical wavevector $\sim$ 5, we have obtained them as 137.2969 and 4.066 respectively for the vertical perturbation 
and $q = 2$, e.g., as provided in TABLE.~\ref{tab:table_cric} which enlists the critical $Re$ 
($Re_{\alpha, cric}$) and the critical wavevector ($k_{z,cric}$) for different rotation parameters. We notice that 
$Re_{\alpha, cric}$ increases as $q$ decreases, as expected from the whole discussion.

Note importantly that the inclusion of rotation does not 
invalidate the Squire theorem which states that a flow that is unstable in 
three dimensions, will be unstable in two dimensions at a lower 
$Re$. It is obvious from 
FIGs.~\ref{fig:Poi_rot_kep_3d_diff_ky_Re_1200} and \ref{fig:eval_cou_poi_re_1500_ky_1_kz_1_9_10_-1_q_kep}. As $k_y$ is 
nonzero and increases further, keeping other parameters fixed, we see that the 
growth rate most unstable mode or least stable mode decreases.

\section{Conclusion}
\label{sec:Conclusion}
In the presence of extra force, plane Couette flow behaves more like plane Poiseuille flow. However, depending on the 
strength of force and the boundary conditions, it may almost behave like plane Couette flow, or the deviation from plane 
Couette flow may be small. Nevertheless, when this flow is studied in the presence of Coriolis effect, it becomes unstable under 
threedimensional perturbations as well as pure vertical perturbations. In fact, rotational effect makes the flow more
unstable and, hence, turbulence inside the underlying shearing box is inevitable.

In the literature before this work, when the hydrodynamic instability of 
the Keplerian accretion flow has been studied in the local region,
the flow has been approximated to plane Couette flow with rotation embedded 
in it. However, recent works (\citealt{Nath_2016,our_work,Raz20}) suggest that the presence
of an extra force (random or constant) is inevitable in such a flow, at least
the effect of external force worth exploring. We, 
therefore, have argued here that the background flow of a local Keplerian
accretion disk will 
deviate from plane Couette flow. We, in fact, have considered here such
deviated background flow modifying to plane Poiseuille flow. This 
modification depends on the strength of force and the boundary conditions. Controlling these two factors, plane Couette flow and also its nature can be 
revived. We know that plane Poiseuille flow is unstable beyond the respective
critical values of certain parameters for planer perturbation. We, therefore, 
can argue that the local Keplerian flow becomes unstable due to the presence of an extra force.

However, the effect of Coriolis force, which is inevitable for shearing box
in the Keplerian
disk, makes the problem more interesting. We know that rotation stabilizes the linear 
shear flow. However, for plane Poiseuille flow, it has opposite effects. 
In the presence of rotation, plane Poiseuille flow 
becomes unstable at a $Re$ which is about two orders of magnitude less than that required for the instability without rotation.
We have shown here that as the rotation parameter $q$ increases, the flow becomes 
more unstable (or at least less stable) for a particular set of 
parameters. The important point here is that the presence of an extra force modifies the local Keplerian flow from linear shear to nonlinear shear 
and the Coriolis effect makes it unstable for a very small $Re$. 
We have also argued that even the presence of a tiny force, that could lead to required amount 
of deviation from 
the linear shear, makes it unstable even in the presence of rotation. Once the flow becomes unstable, eventually it is expected to become 
nonlinear and turbulent. It, therefore, helps us to understand the sub-critical transition to turbulence in hydrodynamic 
accretion flow and other laboratory flows where external forcing, however
tiny be, is unavoidable.

\section*{acknowledgment}
S.G. acknowledges DST India for INSPIRE fellowship. 
We are thankful to 
Laurette S. Tuckerman of the Centre national de la recherche scientifique 
and Dwight Barkley of the University of Warwick for the discussion that influenced us to initiate
this work and further discussion for better presentation. 
We also thank Tushar Mondal and Sudeb Ranjan Datta of Indian Institute of Science for their comments and 
suggestions. We are thankful to the referee for insightful suggestions and comments that help 
present the work in a better way.
This work is partly supported by a fund of Department of Science and Technology (DST-SERB) with research 
Grant No. DSTO/PPH/BMP/1946 (EMR/2017/001226).

\appendix

\section{The derivation of the background flow in the local region of the Keplerian accretion disk}
\label{sec:der_bg_flow_without_force}
Let us consider a fluid element inside the box at the point $P$. With respect to $C$, the flow 
is along the $\phi$ direction. Inside the box, however, the flow will be along the $y$-direction only. Now let us assume 
the velocity at $P$ with respect to the box be $V_Y$. Nevertheless, the velocity at the same point with respect to 
$C$ would be $R\Omega_0 + V_Y$. Had there been no shearing box, the velocity of the fluid at the point $P$ will be 
$\Omega R$ with respect to $C$. Hence,
\begin{eqnarray}
\begin{split}
 R\Omega_0 + V_Y = &\Omega R \Rightarrow V_Y = R(\Omega - \Omega_0)&\\
 = &R (\Omega(R_0+X)-\Omega(R_0)),\ [R-R_0=X<<R_0,R]&\\
 = &R\left[\Omega(R_0) + X \left(\frac{d\Omega}{dR}\right)_{R_0} + ...-\Omega(R_0)\right]&\\
 \cong &RX \left(\frac{d\Omega}{dR}\right)_{R_0}&\\
 = &-q\Omega_0 X \frac{R}{R_0}&\\
 = &-q\Omega_0 X (1+\frac{X}{R_0})&\\
 \cong &-q\Omega_0 X.&
 \label{eq:bg_flow_eq_der}
\end{split}
\end{eqnarray}



\bibliography{mod_bg_flow}{}
\bibliographystyle{aasjournal}



\end{document}